\documentclass[paper]{JHEP}
\usepackage{amsmath}
\usepackage{amssymb}
\usepackage{epsfig}
\usepackage{cite}
\usepackage{feynmf}    % graphic package for constructing Feynman-graphs

\newcounter{hran}
\renewcommand{\thehran}{\arabic{hran}}

\def\bmini{\setcounter{hran}{\value{equation}}
\refstepcounter{hran}\setcounter{equation}{0}
\renewcommand{\theequation}{\thesection.\thehran\alph{equation}}
\begin{eqnarray}}

\def\bminiG#1{\setcounter{hran}{\value{equation}}
\refstepcounter{hran}\setcounter{equation}{-1}
\renewcommand{\theequation}{\thesection.\thehran\alph{equation}}
\refstepcounter{equation}\label{#1}\begin{eqnarray}}

\def\emini{\end{eqnarray}\relax\setcounter{equation}{\value{hran}}\renewcommand{\theequation}{\thesection.\arabic{equation}}}

\def\U{\bbbone}
\def\B{\relax\ifmmode{\mbox{{\rm\bf{B}}}}\else{{\rm\bf{B}}}\fi}
\def\W{\relax\ifmmode{\mbox{{\rm\bf{W}}}}\else{{\rm\bf{W}}}\fi}
\def\X{\relax\ifmmode{\mbox{{\rm\bf{X}}}}\else{{\rm\bf{X}}}\fi}

\def\as{\alpha_{\mbox{\scriptsize s}}}

\def\qq{q\bar{q}}

\def\LQCD{\Lambda_{\mbox{\scriptsize QCD}}}

\def\Re{\mathop{\rm Re}}

\def\bbbone{{\mathchoice {\rm 1\mskip-4mu l} {\rm 1\mskip-4mu l}
{\rm 1\mskip-4.5mu l} {\rm 1\mskip-5mu l}}}
\def\0{{\rm\bf 0}}
\def\sing{{\rm\bf 1}}
\def\s{{\rm\bf s}}
\def\a{{\rm\bf a}}
\def\3{{\rm\bf 3}}
\def\ba3{{\rm\bf \bar{3}}}
\def\b6{{\rm\bf \bar{6}}}
\def\6{{\rm\bf 6}}
\def\8{{\rm\bf 8}}
\def\10{{\rm\bf 10}}
\def\27{{\rm\bf 27}}
\def\U{\bbbone}

\def\al{\alpha}
\def\be{\beta}
\def\sig{\sigma}

\def\de{\delta}

\def\Gam{\Gamma}

\def\cP{{\cal P}}
\def\cQ{{\cal Q}}

\def\cS{{\cal S}}
\def\cN{{\cal N}}
\def\cV{{\cal V}}
\def\cR{{\cal R}}
\def\cM{{\cal{M}}}
\def\cF{{{F}}}

\def\lrang#1{\left\langle#1\right\rangle}

\def\cO#1{{\cal{O}}\left(#1\right)}
\def\half{\mbox{\small $\frac{1}{2}$}}
\def\quart{\mbox{\small $\frac{1}{4}$}}

\def\abs#1{\left|#1\right|}
\def\Tr{{\rm Tr}}
\def\tr{{\rm tr}}
\title{Soft gluons at large angles \\ in hadron collisions}

\author {Yu.L.\ Dokshitzer\footnote{On leave from St.\ Petersburg Nuclear
Institute, Gatchina, St.\ Petersburg 188350, Russia} $\>^a\>$ 
{$\rm and $} 
G.\ Marchesini$\>^{a\,b}$ \\ 
$^a\>$ LPTHE, Universit\'es
Paris-VI-VII, CNRS UMR 7589, Paris, France\\ $\>^b$ Dipartimento di
Fisica, Universit\`a di Milano-Bicocca and \\ $\>\>$ INFN, Sezione di Milano,
Italy}

\abstract{
A general discussion is presented of the single logarithmic soft
factor that appears in two scale QCD observables in processes
involving four partons.  We treat it as the ``fifth form factor'',
accompanying the four collinear singular Sudakov form factors attached
to colliding and outgoing hard partons. The fifth form factor is
expressed in terms of the Casimir operators (squared colour charges)
of irreducible representations in the crossing $t$- and
$u$-channels. As an application we revisit the problem of large angle
radiation in $gg\to gg$ and give a relatively simple solution and
interpretation of the results. We found an unexpected symmetry 
of the soft anomalous dimension under exchange of internal and
external variables of the problem whose existence calls for explanation.}


\preprint{Bicocca--FT--05--19\\
     LPTHE-05-23  \\
     hep-ph/0509078\\
     August 2005}

\begin{document}
\setcounter{footnote}{0}

\begin{fmffile}{hh}
\fmfset{curly_len}{2mm}
\fmfset{wiggly_len}{3mm}

\section{Introduction}

QCD observables that are characterised by two large scales
$Q_0\gg\LQCD$ and $Q\gg Q_0$ are sensitive to multiple gluon radiation
and possess double logarithmic (DL) quark and gluon form factors
depending on the ratio of the scales, $\as\ln^2({Q}/{Q_0})$.  Here $Q$
is the overall hardness scale of the process which is determined by
the underlying parton--parton interaction, and the smaller scale $Q_0$
is introduced by measuring an observable $V$ in specific kinematics.

Hard interaction induces associated emission of relatively soft and/or
quasi-collinear gluons.  We discuss observables $V$ that measure
various characteristics of the {\em secondary parton ensemble}\/ in
inclusive manner. Examples are thrust ($V\!=\!1\!-\!T$) and broadening
($V\!=\!B$) event observables in $e^+e^-$ annihilation and DIS,
accumulated out-of-event-plane momentum in three-jet $e^+e^-$ events
as well as in hard hadron collisions producing jets.

Such observables vanish at the Born level (pure underlying parton
event), $V\!=\!0$, and may reach $V\!=\!\cO{1}$ in the presence of
secondaries that are energetic and non-collinear to primary parton
directions and look as additional jets.  On average,
$\lrang{V}=\cO{\as}$.  Restricting the observable even further, $V\ll
\lrang{V}$, introduces the second scale $Q_0$ setting the maximal
allowed transverse momentum of {\em real}\/ secondary partons. At the
same time, transverse momenta of {\em virtual}\/ gluons are not
bounded.  Break-up of the real--virtual cancellation gives rise then
to the DL form factor suppression of near-to-Born parton
configurations.

DL enhanced corrections originate from emission of soft collinear
gluons. They have simple transparent physical origin, which helps to
analyse and resum them in all orders into exponential Sudakov form
factors attached to each of primary hard partons.

Subleading {\em single logarithmic}\/ (SL) contributions originate
from various sources.  First of all, the DL effects must be treated
with care in order to precisely define the arguments 
% (scale) 
of the DL functions.  Moreover, in certain cases SL corrections emerge due to
``recoil'' that secondary radiation produces upon the primary partons,
which affects determination of the thrust axis or of the event plane.
These SL effects are due to precision treatment of kinematics
(definition of the observable, global momentum conservation, etc.)
and are basically of DL nature.
Moreover, there are direct SL contributions that are suppressed at the
matrix element level and originate either from collinear hard parton
splittings ($z\!\sim\!1$) or from radiation of soft gluons at large
angles. The former is an intrinsic part of jet evolution and is easy
to account for. The latter --- inter-jet radiation --- poses, in
principle, more problems.

In particular, $n$-parton ensembles consisting of energy ordered
gluons radiated at large angles contribute, for example, to particle
energy flow $Q_0=E$ in a given inter-jet direction at the SL level as
$\cO{\as^n\ln^n(Q/Q_0)}$ as was found in~\cite{DS}.  Such ``hedgehog''
multi-gluon configurations are difficult to analyse. The all-order
results for such (so called ``non-global'') observables were obtained
only in the large-$N$ limit~\cite{DS,NONGLOBlargeN}.

{\em Global observables}\/ that acquire contributions from the full
available phase space (rather than from a restricted phase space
``window'' as the non-global ones do) are free from this trouble: only
the hardest among the gluons contributes while the softer ones don't
affect essentially the observable and their contributions cancel
against corresponding virtual terms.  As a result, contribution of
large angle soft gluon radiation reduces to virtual corrections due to
multiple gluons with $k_t>Q_0$ attached to primary hard partons.  They
can be treated iteratively and fully exponentiated, together with DL
terms.

Distributions in various global observables have been resummed in all
orders, with SL accuracy, in the case of underlying QCD processes
involving two or three partons that one finds in $e^+e^-$, DIS and in
hadron--hadron collisions with a hard electro-weak object in the final
state (large-$p_T$ photon, Drell-Yan pair, $Z^0$ boson, etc.).

Carrying out this extensive programme was simplified by the fact that
soft gluon radiation in two- and three-patron systems is essentially
{\em colour-trivial}.  Indeed, let $T_i$ be the colour generator that
enters the amplitude of soft gluon radiation off a hard parton $i$.
Then, due to the colour current conservation ($T_1+T_2=0$ for two and
$T_1+T_2+T_3=0$ for three participating hard partons), the products
$T_iT_j$ that enter the soft gluon radiation probability off the
underlying primary partons are {\em proportional to the identity}\/ in
the colour space.  As a result, the answer could be expressed as the
product of Sudakov {\em form factors}\/ corresponding to each of the
primary partons. Each form factor $F_i(Q_i,Q_0)$ is collinear singular
and given by exponent of the probability of single gluon emission proportional to the
``colour charge'' of the hard parton $i$. Importantly, this answer takes
full care of both DL and SL effects provided one introduces into the
form factors properly defined hardness scales $Q_i$ that depend on the
event geometry.  %~\cite{DKT3}.

The case of four participating partons is the first one when colour
triviality no longer holds~\cite{BS}.  Here two colliding partons can
be found in various colour states.  Radiation of a gluon changes the
colour state of the parton pair and this affects radiation of the next
(softer) gluon. Successive large angle gluon emissions become
interdependent. 
%
%%%%%%%%%%%%%%%%%%%%%%%%%%%%%%%%%%%%%%%%%%%
As a result the product of independent Sudakov form
factors misses essential SL corrections.
The programme of resumming soft SL effects due to large angle gluon
emission in hadron--hadron collisons was addressed in a series of
papers~\cite{BS,KOS,BCMN,ASZ}. It gives an additional form factor
which is not diagonal in colour indices.

This does not constitute a serious technical problem for parton
collisions involving quarks (quark scattering and annihilation into
gluons, QCD Compton) since a $\qq$ pair can be only in two colour
states, {\bf 8}\ and \sing, translating into $\6$ and $\ba3$ for a
$qq$ system.
Gluon--gluon scattering is more involved: here one finds as many as
{\em five}\/ irreducible colour representations: $\8_\a$, $\10 +
\overline{\10}$, \sing, $\8_\s$, \27 ({\em six}\/ in the general
$SU(N)$ case).  The problem of {\em diagonalisation}\/ of the system
of mixing colour channels in $gg$ scattering was formulated by George
Sterman and collaborators in~\cite{KOS}.

In what follows we present a transparent physical interpretation of
large angle radiation effects which can be expressed in terms of the
``{\em fifth form factor}\/'' that depends on charge exchange in the
cross ($t$- and $u$-) channels of the scattering process. We have
summarized part of the results in \cite{DMlett}. With account of
logarithmically enhanced, DL and SL, virtual corrections the matrix
element $M_0$ becomes
\begin{equation}
\label{eq:fiveff}
 M_0 \>\to\>\prod_{i=1}^4 F_i(Q_i,Q_0) \cdot \cF_X(\tau_0)\cdot M_0
 \>\equiv\> \prod_{i=1}^4 F_i(Q_i,Q_0) \cdot M(\tau_0).
\end{equation}  
The fifth form factor $\cF_X$ (where ``$X$'' stands for
``cross-channel'') is collinear finite and therefore a SL function
depending on the logarithmic variable
\begin{equation}
  \label{eq:tau}
 \tau_0 = \int_{Q_0}^{Q}\frac{dk_t}{k_t}\frac{\as(k_t)}{\pi}\,.
\end{equation}
It also depends on kinematical variables $s,t,u$ of the hard process.

The fifth form factor exists in the QED context as well, but the
specificity of QCD scattering is that here it has
essentially non-Abelian structure since its exponent contains
non-commuting $t$- and $u$-channel (squared) colour charges.  It
reduces, however, to the Abelian construction in special case of small
angle scattering, $|t|\ll s$ or $|u|\ll s$, where it describes
reggeization of corresponding cross channel exchange states (in
particular, gluon and quark reggeization).

We also remark upon the existence of a strange symmetry between the
external (scattering angle) and internal variables (rank of the gauge
group). Elucidating the nature of this unexpected symmetry calls for
an effort on the part of the ``theoretical-theory'' community.

It is clear that the programme of resumming soft SL effects due to
large angle gluon emission requires special care to be taken of
collinear enhanced DL contributions that have to be treated with
subleading SL accuracy. This involves precise definition of the
arguments $Q_i$ of the DL form factors and of the parton densities, an
accurate account of running coupling effects, employing Mellin/Fourier
transformation to carefully factorise multiple emissions, etc.
Carrying out this programme results in the expression for the
(integrated) distribution $\Sigma(Q_0,Q)$ that has the following
general form:
\begin{equation}
\label{eq:Sig}
   \Sigma(Q_0,Q) \>=\> \Sigma^{{\mbox{\scriptsize coll}}}(Q_0,Q) \cdot
   \cS_X(\tau_0).
\end{equation}
Here the ``collinear factor'' $\Sigma^{{\mbox{\scriptsize coll}}}$
originates from the four Sudakov form factors in \eqref{eq:fiveff} and
embodies the parton densities, while the SL soft factor $\cS_X$ is due
to the cross-channel ``fifth form factor'' $\cF_X$,
\begin{equation}\label{eq:cS}
    \cS_X(\tau_0) \>=\> \frac{\Tr (M^\dagger(\tau_0)\>M(\tau_0))}{\Tr
      (M_0^\dagger\> M_0)}, \quad M(\tau_0)= \cF_X(\tau_0)\cdot M_0\,.
\end{equation}
The paper is organised as follows. In the next Section we give the
general treatment of soft gluon radiation accompanying hard scattering
of arbitrary colour objects. We elucidate the origin of the soft SL
factor (``fifth form factor'') as due to large angle radiation which
is governed by colour being exchanged in crossing channels. We
describe the general procedure for calculating the corresponding soft
anomalous dimension in terms of Casimir operators of $t$- and
$u$-channel colour states. We also demonstrate cancellation of the
divergent piece of non-Abelian Coulomb phase.

Section~3 is devoted to gluon--gluon scattering. Here we give an
ergonomic solution for eigenvalues and eigenvectors of the anomalous
dimension matrix and present the answer for the soft factor in various
special cases.

The main ingredients of the paper are collected in Conclusions where
in particular we draw reader's attention to unexpected symmetry of the
problem.

\section{Soft gluons and the fifth form factor}

Consider a hard process involving $N_p$ primary partons.  Let $T_i^b$
be the colour generator that enters the amplitude of emission of a
soft gluon with momentum $k^\mu=(\omega, {\bf k})$ off a hard parton
$i$. The emission is described by the current
\begin{eqnarray}
\label{eq:jdef}
   j^{\mu,b}(k) = \sum_{i=1}^{N_p} \frac{\omega \, p_i^\mu}{(k
     p_i)}T_i^b\,; \qquad
    \sum_{i=1}^{N_p} T_i^b \>=\> 0.
\end{eqnarray}
The latter equation represents conservation of the colour charge which
guarantees current conservation,   $k_\mu j^\mu=0$.
Squaring the current to compute the distribution we obtain
\begin{equation}
 \label{eq:j2gen}
 -j^2(k) \>=\> -2\sum_{i>j} T^b_iT^b_j \cdot w_{ij}(k), 
\end{equation} 
where $w_{ij}$ is the ``dipole antenna'' distribution depending on
parton angles, 
\begin{equation}\label{eq:wdef}
\begin{split}
 w_{ij} &= \frac{\omega^2\,(p_ip_j)}{(kp_i)(kp_j)} =
 \frac{\xi_{ij}}{\xi_i\xi_j}, \qquad  
  \xi_{ij} = 1-\cos\Theta_{{\bf p}_i, {\bf p}_j}, \>\> 
   \xi_i = 1-\cos\Theta_{{\bf p}_i, {\bf k}}.
\end{split} 
\end{equation}

 \subsection{Colour triviality of the $N_p=2,3$ cases and Sudakov form
   factors}

For $N_p=2,3$ the products of colour generators that enter
\eqref{eq:j2gen} are {\em proportional to the identity}\/ in the
colour space.  Indeed, in the case of two quark jets in $e^+e^-$ there
is one dipole only and the corresponding colour factor is simply
$$
-2T_1^bT_2^b = T_1^2+T_2^2 = 2C_F.
$$
For $N_p\!=\!3$ (three jet $\qq g$ events) all colour factors can also
be expressed as combinations of the Casimir operators $T_i^2$ since by
virtue of the charge conservation in~\eqref{eq:jdef}
$$
-2T_1T_2\>=\> T_1^2 + T_2^2 - T_3^2 \qquad \mbox{(and cyclic)},
$$
and one can write
\begin{equation} \label{eq:j2qqg}
\begin{split}
-j^2(k) &\>=\> T_1^2\cdot W^{(1)}_{23}(k) + T_2^2\cdot W^{(2)}_{13}(k)
+ T_3^2 \cdot W^{(3)}_{12}(k)\,,
\end{split}
\end{equation} 
 where we have introduced the dipole combinations 
\begin{equation}\label{eq:W123def}
   W^{(1)}_{23} = w_{12}+w_{13}-w_{23}\,.
\end{equation}
The essential property of the distribution \eqref{eq:W123def} is that
it is collinear singular {\em only}\/ when ${\bf k} \parallel {\bf
  p}_1$. This singularity contributes proportional to the
corresponding Casimir operator, in accord with general factorisation
property.
Integrating \eqref{eq:W123def} over angles gives
\begin{equation}\label{eq:Wangint}
\int\frac{d\Omega}{4\pi}\> W ^{(1)}_{23} \>=\>
   \ln\frac{(p_1p_2)(p_1p_3)}{(p_2p_3)\, m^2} =
   \ln\frac{p_{t1,23}^2}{2\,m^2}\,,
\end{equation}
with $p_{t1,23}$ transverse momentum of parton $p_1$ in the cms of the
$(p_2,p_3)$ pair, and $m^2$ the collinear cutoff. Exponentiating the soft
gluon current squared \eqref{eq:j2qqg} leads to the product of three
Sudakov form factors. The collinear cutoff $m$ disappears when the
virtual and real contributions (to a collinear and infrared safe
observable) are taken together, and gets replaced by the proper
observable dependent scale $\cO{Q_0}$, see examples in~\cite{QVscale}.

For the system of $q(1)$, $\bar{q}(2)$ and a hard gluon $g(3)$ we have
$T_1^2=T_2^2=C_F$, $T_3^2=C_A$. Applying \eqref{eq:Wangint} to the
full emission probability \eqref{eq:j2qqg},
\begin{equation}
\begin{split}
-\int\frac{d\Omega}{4\pi}\,j^2(k) &=
% C_F\ln\frac{(p_1p_2)(p_1p_3)}{(p_2p_3) m^2} +
% C_F\ln\frac{(p_2p_1)(p_2p_3)}{(p_1p_3) m^2} +
% N\ln\frac{(p_3p_1)(p_3p_2)}{(p_1p_2) m^2} \\
 2C_F \ln\frac{s_{\qq}}{2m^2} + N \ln\frac{p_{t3}^2}{2 m^2}\, ,
\end{split}
\end{equation}
we see that the answer for the $\qq g$ system can be represented as
the product of the two quark form factors at the scale
$s_{\qq}=2(p_1p_2)$ (cms energy of the $\qq$ pair) and the gluon form
factor taken at the scale $p_{t3}^2=2(p_3p_1)(p_3p_2)/(p_1p_2)$
(invariant gluon transverse momentum)~\cite{DKT3}.

\subsection{Large angle gluon radiation and the
  fifth form factor}
 
Now we turn to processes involving four hard partons and consider
$1+2\to 3+4$ scattering characterised by kinematical variables
$$
  s=2p_1p_2,\quad -t=2p_1p_3, \quad -u=2p_1p_4 \,.
$$
In what follows we will treat all three Mandelstam invariants as being
of the same order of magnitude and discuss small angle scattering
separately.

Soft gluon radiation off the four-parton ensemble is given by
\eqref{eq:jdef} for $N_p\!=\!4$, with $T_i^b$ quark, antiquark or
gluon generator depending on the nature of participating parton $i$.
We remark that within the convention \eqref{eq:jdef} the generators
are taken as if all partons were {\em incoming}\/ (e.g., the actual
colour charges of the outgoing partons 3 and 4 equal $-T_3^b$ and
$-T^b_4$).

It is straightforward to verify that the sum of dipoles in
\eqref{eq:j2gen} can be identically represented as
\begin{equation}
\label{eq:alter}
\begin{split}
  - j^2(k) =\> & T_1^2\, W^{(1)}_{34}(k) + T_2^2\, W^{(2)}_{34}(k) +
  T_3^2\, W^{(3)}_{12}(k) +  T_4^2\, W^{(4)}_{12}(k)  \\
  &+\> T_t^2\cdot A_{t}(k) \> \>+ T_u^2\cdot A_u(k)\,.
\end{split}
\end{equation} 
The first four terms form the product of the form factors attached to
participating hard partons as before. Their angular integrals, see
\eqref{eq:Wangint}, are the same and give
\begin{equation}
\label{eq:Qtudef}
 \ln\frac{Q^2}{2m^2},\quad   \mbox{with}\>\>  Q^2=  \frac{tu}{s} \>=\>
 s\,\sin^2\Theta_s,
\end{equation}
which combination of invariants becomes the hard scale of the process
common for all four Sudakov factors.

The last two terms in \eqref{eq:alter} give rise to the {\em fifth
  form factor}\/ as discussed in the Introduction.  The two operators
$T_t^2$ and $T_u^2$ are the squared colour charges exchanged in the
$t$ and $u$ channels of the scattering process,
\begin{equation}
\label{eq:TT}
 T_t^2 = (T_3+T_1)^2 = (T_2+T_4)^2, \quad T_u^2= (T_4+T_1)^2= (T_2+T_3)^2.
\end{equation} 
They do not commute and this is where the call for the colour
diagonalisation programme comes from.  The new angular dipole
combinations that accompany these operators are given by
\begin{equation}
\label{eq:AtAu}
A_t= w_{12} +w_{34}-w_{13}-w_{24}\,,\quad 
A_u= w_{12} +w_{34}-w_{14}-w_{23}\,,
\end{equation} 
and, unlike the dipoles $W^{(i)}_{jk}$, are integrable in angles:
\begin{equation}
\label{eq:AtAuangint}
\int\frac{d\Omega}{4\pi}\, A_t(k) \>=\>  2\ln\frac{s}{-t}\,; \qquad 
\int\frac{d\Omega}{4\pi}\, A_u(k) \>=\>  2\ln\frac{s}{-u}\,.
\end{equation} 
Contrary to the first four DL contributions (that give rise to Sudakov
form factors), the additional contribution originates from coherent
gluon radiation at angles {\em larger than the cms scattering angle
  $\Theta_s$}.
Indeed, in the cms of colliding partons
$(\xi_1+\xi_2=\xi_{12}=\xi_{34}=\xi_3+\xi_4=2)$ we have
\begin{equation}
 \begin{split}
   A_t =& 
%\left[\frac{1}{a_1} +\frac{1}{a_2}\right] +
%   \left[\frac{1}{a_3} +\frac{1}{a_4}\right]
%   -  \frac{a_{13}}{a_1a_3}  - \frac{a_{24}}{a_2a_4} \\
%%    &=  \frac{a_{12}}{a_1a_2} + \frac1{a_1}\frac{a_1-a_{13}}{a_3} 
%%   + \frac1{a_2}\frac{a_2-a_{24}}{a_4}  \\
 \frac1{\xi_1} \left[1- \frac{\xi_{13}-\xi_1}{\xi_3} \right] +
 \frac1{\xi_2} \left[1- \frac{\xi_{24}-\xi_2}{\xi_4} \right].
 \end{split}
\end{equation}
Upon integration over the azimuth angle $\phi$ of the radiated gluon
around the direction $z$ of colliding parton momenta ${\bf p}_1=-{\bf
  p}_2$,
\begin{equation}
\label{eq:largeang}
 \begin{split} 
  \int\frac{d\phi}{2\pi} \> A_t(k)
%   = \frac1{\xi_1} \left[1- \frac{\xi_{13}-\xi_1}{|\xi_{13}-\xi_1|} \right] +
%   \frac1{\xi_2} \left[1- \frac{\xi_{24}-\xi_2}{|\xi_{24}-\xi_2|} \right] 
   = \frac1{\xi_1}\vartheta(\xi_1-\xi_{13}) +
   \frac1{\xi_2}\vartheta(\xi_2-\xi_{24}).
 \end{split}
\end{equation}
% where we used $\xi_{24}=\xi_{13}$. 
Thus, the gluon emission angle is limited {\em from below}\/ by the
$t$-channel scattering angle, $\xi_1>\xi_{13}=\xi_{24}$.  Analogously,
in the $T_u^2$ term the lower limit is given by the $u$-channel
scattering angle, $\xi_1>\xi_{14}$.  Polar angle integration then
gives the logarithms of the ratio of Mandelstam invariants as stated
in \eqref{eq:AtAuangint}.

The r\^ole of coherent large angle gluon radiation driven by
$t$-channel colour exchange was elucidated in~\cite{Higgs} where
comparison was made of the distributions of hadrons accompanying
production of Higgs bosons via gluon--gluon and $W^+W^-$ fusion.

 \subsection{Analysis of virtual corrections}

Each parton channel has various colour channels:
\begin{equation}
  \label{eq:a1234}
 a_1+a_2\>\to\> a_3+a_4\,.
\end{equation}
The hard scattering matrix element for a given channel is a function
of four colour indices $\{a_i\}$ of participating partons: $a=1,\ldots
N$ for quark and $a=1,\ldots N^2\!-\!1$ for gluon:
\begin{equation}
 (M_0)^{a_1a_3}_{ a_2 a_4}\,; \quad \sig_0 \>\propto\> \sum_{a_i}
 (M_0^\dagger)^{a_3 a_1}_{a_4 a_2}\> (M_0)^{a_1 a_3}_{a_2 a_4} \> =\>
 \Tr\left( M_0^{\dagger}\cdot M_0\right) .
\end{equation}
To address the problem of all-order analysis of soft radiation with
single logarithmic accuracy, it suffices to study virtual corrections
to hard patron matrix element $M$ due to multiple soft gluons.  Then,
real production cross sections can be obtained simply by ``cutting''
the product $M^\dagger \cdot M$.  Let $k_V$ be the contribution to the
observable $V$ of the soft (real) gluon $k$. In the integration region
$0<k_V \ll Q_0$ positive contribution to the cross section due to real
production of gluons whose effect upon the observable is negligibly
small cancels against negative virtual correction. After this standard
real--virtual cancellation, the resulting distribution is determined
by the virtual factor originating from the complementary integration
region $Q_0< k_V < Q$.
 
The soft cross channel form factor $\cF_X(\tau_0,\{p_i\})$, the fifth
form factor in \eqref{eq:fiveff}, is expressed in terms of the matrix
$M(\tau_0)$ that is obtained by considering virtual corrections to
$M_0$ due to soft gluons with $k_t>Q_0$:
\begin{equation}
  \label{eq:cF} 
  M_0 \> \Longrightarrow \> M(\tau_0) = \cF_X(\tau_0)
  \cdot M_0\,, \quad \cF_X(0)=1\,.
\end{equation}
For $Q_0\sim Q$ we have $\tau_0\ll1$ and the matrix $M(\tau_0)$
reduces to $M_0$.  The cross section then acquires the ``soft factor''
$\cS_X(\tau_0)$ given in \eqref{eq:cS} and we may write
\begin{equation}
\cF_X^\dagger(\tau_0) \cdot \cF_X(\tau_0) \quad \Longrightarrow \quad 
 \cS_X(\tau_0) ,
\end{equation}
where $\cF_X$ is a matrix in the colour space that will be dealt with
in what follows.

The $\tau$ dependence of $M(\tau)$ can be extracted using the
differential equation~\cite{BS} that arises from the following
iterative procedure.  By virtue of soft gluon factorisation, the {\em
softest}\/ gluon $k$ is emitted from the four external primary parton
lines.  Differentiating over its transverse momentum we obtain
\begin{equation}
  \label{eq:eveq}
  \partial_{\tau}\,M(\tau) \>=\>   G(\tau)\cdot M(\tau)\,, \quad
  \tau = \int_{k_t}^{Q}\frac{dk'_t}{k'_t}\frac{\as(k'_t)}{\pi}\,,
\end{equation}
where $G(\tau)$ multiplies the matrix element $M(\tau)$ dressed by
gluons that are {\em harder}\/ than $k_t$.  The soft anomalous
dimension $G$ is a colour matrix and is a function of $s$, $t$ and
$u$.  It is not symmetric and is complex due to the $s$-channel gluon
exchanges (Coulomb phases).

Soft virtual corrections to the hard scattering matrix element can be
split into two pieces: eikonal and Coulomb contributions.

\paragraph{Eikonal contribution.}
The first virtual contribution equals {\em minus one half}\/ of the
eikonal current squared \eqref{eq:alter} that we considered above in
the context of {\em real}\/ soft gluon radiation, and cancels it in
the part of the phase space that is open for real gluons: $k_V\sim k_t
\sim \omega <Q_0$. The colour trivial collinear logarithmic pieces in
\eqref{eq:alter} are extracted and included into the exponents of four
Sudakov form factors $F_i$ at the hard scale $Q$ given in
\eqref{eq:Qtudef}.  The remaining soft SL cross channel contributions,
upon integration over gluon angles, give the soft anomalous dimension
responsible for virtual suppression coming from the region $Q_0 < k_t
< Q$ as stated in \eqref{eq:tau}:
\begin{equation}
  \label{eq:Gam-real}
 G^{\rm eik} = G_{\rm real} + G^{\rm eik}_{\rm virt} \> =\>
 - \left( T_t^2\cdot \ln\frac{s}{-t}
 + T_u^2 \cdot \ln\frac{s}{-u} \right), \quad Q_0< k_t <  Q\,.
\end{equation}
Let us stress that this matrix is a constant in $\tau$, the
$\tau$-dependence entering through the boundary in $k_t$.

\paragraph{Coulomb gluons.} 
An additional virtual contribution arises when a soft virtual gluon
connects two incoming or two outgoing partons.  While in the eikonal
contributions (both real and virtual) it was the gluon line that was
put on-shell, the Coulomb correction is obtained by putting on-shell
the two hard parton lines in the intermediate state. This contribution
can be extracted by considering in \eqref{eq:j2gen} only the
$(p_1,p_2)$ and $(p_3,p_4)$ interactions and replacing $w_{12}$ and
$w_{34}$ by $i\pi$. The imaginary contributions present in the
diagonal (Sudakov) pieces in \eqref{eq:alter} give rise to Abelian
Coulomb phase which fully cancels upon multiplication of the amplitude
by the conjugate one.  Imaginary part of the soft 
% non-diagonal 
anomalous dimension reads
\begin{equation}
  \label{eq:Gam-C}
  G_{\rm C} = i\pi \left(T_t^2 + T_u^2\right), \qquad 0< k_t <Q;
\end{equation}
it provides the amplitude with a non-Abelian Coulomb phase factor. 

Combining real emission, virtual eikonal and Coulomb corrections, the
final result for the soft anomalous dimension $G$ becomes
\begin{eqnarray}
 \label{eq:Gam-fine} 
 G(\tau) &=& \Gamma\cdot\vartheta(\tau_0-\tau),
 \qquad \Gamma\>\equiv\> G^{\rm eik}+ G_{\rm C} = -(T^2_t\, T +
 T^2_u\, U) \,;\\ \label{eq:TU} T &=& \ln\frac{s}{-t} - i\pi\,,
\quad U  = \ln\frac{s}{-u} - i\pi\,,
\end{eqnarray}
for $Q_0 < k_t < Q$, and 
\begin{equation}
 \label{eq:GC-fine}
\begin{split}
  G(\tau) &= \Gamma_C\cdot\vartheta(\tau-\tau_0),
\qquad \Gamma_C  \equiv G_{\rm C}  
 = i\pi\left(T^2_t + T^2_u\right), \qquad \>\> { }
\end{split}
\end{equation}
for $0 < k_t < Q_0$.

\paragraph{Exponentiation.}

The evolution equation \eqref{eq:eveq} has to be integrated over
$\tau$ from $0$ ($k_t\!=\!Q$) up to $\Lambda\to \infty$
($k_t\!=\!0$). The formal solution is given by a $\tau$-ordered
exponent
\begin{equation}
 M(\tau_0) \>=\> P_{\tau}\exp \left\{ \int_0^{\Lambda}
 d\tau\,G(\tau)\right\}\cdot M_0\,.
\end{equation}
Since $G(\tau)$ assumes (different) {\em constant}\/ values for
$0<\tau<\tau_0$ and $\tau_0<\tau<\infty$, using
\eqref{eq:Gam-fine}--\eqref{eq:GC-fine} we obtain
\begin{equation}
\label{eq:qui}
  M(\tau_0) \>=\> e^{(\Lambda-\tau_0)\Gam_C} \, e^{\tau_0\,\Gam}\cdot
  M_0\,.
\end{equation}

\paragraph{Cancellation of divergent Coulomb phase.} 

For $\Lambda\to\infty$ ($k_t\to0$) the first factor in \eqref{eq:qui}
diverges.  However, since exchanging a gluon between two incoming (or
outgoing) partons obviously does not affect their overall colour
state, the sum of the two matrices $T^2_t+T^2_u$
is necessarily {\em diagonal}\/ in $s$-channel colour,
\begin{equation}
T_t^2 + T_u^2 \>=\> - T_s^2 + \sum_{i=1}^4 T_i^2 \,, \qquad T_s=T_1+T_2=-(T_3+T_4)\,.
\end{equation}
Therefore the Coulomb matrix $\Gamma_C$ in \eqref{eq:GC-fine} is anti-Hermitian.  
Thus the first factor in \eqref{eq:qui} becomes a
unitary matrix and cancels upon multiplication with the conjugate
amplitude in \eqref{eq:cS}.  Therefore, in the calculation of physical
distributions one can effectively neglect the divergent Coulomb phase
factor in \eqref{eq:qui} and use
\begin{equation}
  \label{eq:M-fine}
  M(\tau_0) \>=\> e^{\tau_0\,\Gam}\cdot M(0)\,,\qquad 
  \Gam= -\left(T^2_t\cdot T +  T^2_u\cdot U\right).
\end{equation}
Let us stress that cancellation of the non-Abelian Coulomb phase is
only partial.  An imaginary Coulomb contribution coming from the
virtual gluon momentum region $Q_0<k_t<Q$ is still present and enters
$\Gamma$ through complex logarithms $T$ and $U$ defined
in~\eqref{eq:TU}.

\subsection{Colour structure}

To evaluate the matrices $T_t^2$ and $T_u^2$ in \eqref{eq:M-fine} we
turn to the analysis of the colour structure of the process in various
channels.

The operator $T_t^2$ is a number in a given colour state of the
$t$-channel parton pair, $(p_1,p_3)$ and/or $(p_2,p_4)$. Therefore, in
the $t$-channel projector basis it is the diagonal matrix of Casimirs.
Similarly, $T_u^2$ is diagonal in the $u$-channel projector basis
$(p_1,p_4)$ or $(p_2,p_3)$.

For calculation of the $s$-channel observables it is natural, however,
to describe colour states from the $s$-channel point of view.  We will
use the $s$-channel basis of projectors $\cP_{\al}$ onto irreducible
$SU(N)$ representations that are present in the colour space of two
incoming partons $(p_1,p_2)$ or, equivalently, outgoing
$(p_3,p_4)$. The completeness relation reads
\begin{equation}
  \label{eq:Unity}
\bigl(\U\bigr)_{a_2a_4}^{a_1a_3} = 
\delta^{a_1,a_3}\delta_{a_2,a_4}\,,\qquad
\U=\sum_{\al=1}^{\cN}\cP_{\al}\,, \qquad 
\Tr(\cP_{\al}) = K_{\al}\,.
\end{equation}
Here $K_{\al}$ is the dimension of the representation $\al$, and $\cN$
the number of irreducible representations involved.  The sum over all
$K_{\al}$ equals the total number of colours states of the system of
two incoming (or outgoing) partons.  The matrix element can be
expressed as a sum over colour projectors 
\begin{equation}
\label{eq:M-tau} M_0 =\sum_{\al=1}^{\cN}m_{0\al}\,\cP_{\al}\,, \qquad
M(\tau)=\sum_{\al=1}^{\cN}m_{\al}(\tau)\,\cP_{\al}\,.  
\end{equation}
In order to compute $T_{t,u}^2$ we need to know the transition
matrices connecting $s$- and $t$-/$u$-channels projectors.
Introducing
\bminiG{eq:Kall}
\label{eq:Kts}
  \cP^{(t)} &=& K_{ts} \cdot \cP\,, \quad \cP  = K_{st}\cdot
  \cP^{(t)}\,;  \qquad K_{st}=(K_{ts})^{-1}; \\ 
\label{eq:Kus}
  \cP^{(u)} &=& K_{us} \cdot \cP, \quad \cP  = K_{su}\cdot
  \cP^{(u)};  \qquad K_{su}=(K_{us})^{-1},
\emini
we have
\begin{equation}
\label{eq:Trepro}
 \left(T_t^2\right)_{\al\be} = \sum_\rho (K_{st})_{\al\rho}\,
 c^{(t)}_\rho \, (K_{ts})_{\rho\be} \,,
\end{equation}
where the indices $\al$ and $\be$ mark irreducible representations of
the colour group of the pair of incoming partons ($s$-channel) and
$\rho$ --- representations of the $t$-channel pair. In particular,
$c^{(t)}_\rho$ stand for the Casimir operator of the $t$-channel
representation $\rho$.

In general, $K$ are not necessarily square matrices. For example, for
the process $\qq\to gg$ we have two colour states in the $s$-channel,
\sing\ and \8, while there are three in the $t$-channel: \3, ${\bf
  \bar{6}}$ and {\bf 15}.

We conclude this section devoted to general discussion of the physics
and technical ingredients of the analysis of the fifth form factor by
presenting the final general expression which holds for scattering of
arbitrary colour objects:
\begin{equation}
  \label{eq:cSans}
    \cF_X(\tau_0) \>=\> \exp\left\{ -\tau_0  \left( T^2_t\cdot T +
        T^2_u\cdot U\right)\right\},
\end{equation}
where the cross-channel squared colour charge matrices $T_t^2$ and
$T_u^2$ have to be computed using \eqref{eq:Trepro}.  We will
demonstrate this computation on the concrete case of gluon--gluon
scattering.

\section{Gluon--gluon scattering}

The $gg\to gg$ amplitude has $(N^2\!-\!1)^4$ colour indices. We start
by introducing the $s$-channel projector basis for this process.

\subsection{Irreducible representations and projectors}

The colour state of two gluons can be characterised in terms of
irreducible representations. In $SU(3)$ we have
\begin{equation}
  \label{eq:irred3}
  \bf{glue} \otimes \bf{glue} 
  = \bf{8}_\a + \10 + \sing + \bf{8}_\s + \27,
\end{equation}
where $\bf{8}_a$ and $\10$ mark {\em antisymmetric}\/ representations
(octet and the sum of the decuplet and anti-decuplet) and three {\em symmetric}\/ ones are
the singlet ($\sing$), octet ($\bf{8}_s$) and the high symmetric
tensor representation ($\27$) with corresponding dimension.  
In the general case of $SU(N)$ (with $N>3$) we have an additional 
{\em symmetric}\/ representation (which we mark \0):
\begin{equation}
  \label{eq:irredN}
  \bf{glue} \otimes \bf{glue} 
  = \bf{8}_\a + \10 + \sing + \bf{8}_\s + \27 +\0.
\end{equation}
Thus we will keep using the $SU(3)$ motivated names in spite of the
fact that the dimensions of corresponding representations are actually
different from $8$, $2\times 10$, etc.:
\begin{equation}
\label{eq:Fs}
\begin{split}  
  K_\sing &= 1, \qquad  K_\a     = K_\s = N^2-1, \qquad
  K_{\10} = 2\times \frac{(N^2-1)(N^2-4)}{4}, \\
  K_{\27} &= \frac{N^2(N-1)(N+3)}{4}, \qquad\qquad \quad
  K_{\0}  \>=  \frac{N^2(N+1)(N-3)}{4}.
\end{split}
\end{equation}
(Mark that for $N=3$ we have indeed $K_\0=0$.)  It is straightforward
to check that the sum of the dimensions \eqref{eq:Fs} gives
$(N^2-1)^2$ as expected.

As was explained above, it is convenient to work in the colour space
of hard gluon scattering in the $s$-channel.  We construct the basis
of $s$-channel projectors $\cP_\al$ ordered as follows:
\begin{equation} \label{eq:basis}
  \al=\{\8_\a,\10,\sing,\8_\s,\27,\0\}\,.
\end{equation}
The projectors are explicitly constructed in Appendix~\ref{AppProj}.
They satisfy the completeness relation
\begin{eqnarray}
  \label{eq:projsum}  
 \sum_{\al=1}^6\cP_{\al} &=& \cP_\a + \cP_{\10} + \cP_{\sing} +
 \cP_\s + \cP_{\27} + \cP_{\0} = \U\,; \quad 
 \bigl(\U\bigr)_{a_2a_4}^{a_1a_3} =  \de^{a_1,a_3}\de_{a_2,a_4}\,. 
\end{eqnarray}
Dimension of a given representation \eqref{eq:Fs} can be calculated by
taking trace of the corresponding projector:
\begin{eqnarray} \label{eq:trace}  
 \Tr\,(\cP_\al) &\equiv& \sum_{a_1,a_2}
 \left(\cP_\al\right)^{a_1\,a_1}_{a_2\,a_2} \>=\> K_\al\,.
\end{eqnarray}
Casimir operators of all six representations are calculated in
Appendix~\ref{AppSEx}.  In our basis \eqref{eq:basis} the diagonal
matrix of Casimir operators reads
\begin{equation}
\label{eq:C2matr}
(T^a)^2_{\al\be} \>=\> (C_2)_{\al\be} = \delta_{\al\be} \cdot c_\al
\,, \qquad c_\al=
\{N,\>2N,\>0,\>N,\>2(N\!+\!1), \>2(N\!-\!1)\}.
\end{equation}
This matrix enters the expression \eqref{eq:Trepro} for the anomalous
dimension $\Gamma$.  Matrices $K_{ts}$ and $K_{us}$ that rotate the
$s$-channel projector basis into $t$- and $u$-channels are calculated
in Appendix~\ref{AppRepro}.

\subsection{Diagonalisation of the anomalous dimension matrix}

We shall represent the anomalous dimension matrix as
\begin{equation}
 \Gamma \>=\> - N(T+U)\cdot \cQ\,,
\end{equation}
where the matrix $\cQ$ depends on the ratio of the logarithmic
variables
\begin{equation}\label{eq:bdef}
  b \>\equiv\> \frac{T-U}{T+U}\,.
\end{equation} 
\begin{eqnarray}
\label{eq:Qnew}
 \cQ &=& \left( \begin{array}{rrrrrr} \frac32 & 0 & \>\> -2b & \>\>
 -\frac{1}{2} b & -\frac{2}{N^2}b & -\frac{2}{N^2}b \\[2mm] 0 & 1 & 0
 & -b & \>\> -\frac{(N+1)(N-2)}{N^2}b & \>\> -\frac{(N-1)(N+2)}{N^2}b
 \\[2mm] -\frac{2}{N^2-1}b & 0 & 2 & 0 & 0 & 0 \\[2mm] -\frac12 b&
 -\frac{2}{N^2-4}b & 0 & \frac32 & 0 & 0 \\[2mm] -\frac{N+3}{2(N+1)}b
 & \>\> -\frac{N+3}{2(N+2)}b & 0 & 0 & \frac{N-1}{N} &0\\[2mm]
 -\frac{N-3}{2(N-1)}b & -\frac{N-3}{2(N-2)}b & 0 & 0 & 0 &
 \frac{N+1}{N} \end{array} \right)
\end{eqnarray}
It is worth observing that the matrix elements of the states \27 and
\0 (two last rows and columns) are formally related by the operation
$N\to -N$.

\subsubsection{Eigenvalues and eigenvectors}

Six eigenstates of $\cQ$ naturally split into two groups of three.

\paragraph{The first three.}
The first three eigenvalues are $N$-independent:   
\begin{equation}
  \label{eq:E123}
   E_1 = 1\,, \qquad
   E_2 = \frac{3-b}2\,, \qquad
   E_3 = \frac{3+b}2\,.
\end{equation}
The eigenvectors $\cV_1,\cV_2,\cV_3$ corresponding to these
eigenvalues  are
\begin{equation}
\label{eq:V123}
{\cal{V}}_{1,2,3} \!=\!  \left[
\begin{array}{c}
 \displaystyle 0  \\[2ex]
 \displaystyle 1  \\[2ex]
 \displaystyle 0 \\[2ex]
 \displaystyle \frac{4b}{N^2\!-\!4}  \\[2ex]
 \displaystyle -\frac{b\,N(N\!+\!3)}{2(N\!+\!2)} \\[2ex]
 \displaystyle \frac{b\,N(N\!-\!3)}{2(N\!-\!2)}
\end{array}
 \right]
\quad
\left[
\begin{array}{c}
 \displaystyle 1\!+\!b  \\[2ex]
 \displaystyle -2b  \\[2ex]
 \displaystyle \frac{4b}{N^2\!-\!1} \\[2ex]
 \displaystyle 1+ \frac{b\,(N^2\!-\!12)}{N^2\!-\!4}  \\[2ex]
 \displaystyle -\frac{b\,N(N\!+\!3)}{(N\!+\!1)(N\!+\!2)}\\[2ex]
 \displaystyle -\frac{b\,N(N\!-\!3)}{(N\!-\!1)(N\!-\!2)} 
\end{array}
 \right]
\quad
\left[ 
\begin{array}{c}
 \displaystyle -1\!+\!b  \\[2ex]
 \displaystyle -2b  \\[2ex]
 \displaystyle  -\frac{4b}{N^2\!-\!1} \\[2ex]
 \displaystyle  1- \frac{b\,(N^2\!-\!12)}{N^2\!-\!4} \\[2ex]
 \displaystyle \frac{b\,N(N\!+\!3)}{(N\!+\!1)(N\!+\!2)} \\[2ex]
 \displaystyle \frac{b\,N(N\!-\!3)}{(N\!-\!1)(N\!-\!2)}
\end{array}
 \right]. 
\end{equation}
The states 2 and 3 are related by the crossing transformation
$t\leftrightarrow u$. In particular, the eigenvector $\cV_3$ is
obtained from $\cV_2$ by $b\to -b$ and changing the sign of the
antisymmetric projector components (first two rows).

We also remark that $N$ enters the $\cP_{\27}$ and $\cP_\0$ components
with the opposite sign thus reflecting the symmetry of the matrix
elements of $\cQ$.

\paragraph{The last three.}
The last three eigenvalues solve the cubic equation 
\begin{equation}
\label{eq:E13form}
\left[E_i-\!\frac{4}{3}\right]^3 
- \frac{(1+3b^2)(1+3x^2)}{3}\left[E_i-\!\frac43\right] 
- \frac{2(1-9b^2)(1-9x^2)}{27}  \>=\> 0, 
\end{equation}
where we have introduced the notation
\begin{equation}
  \label{eq:1N}  x \>=\> \frac1N\,.
\end{equation}
Solutions can be parametrised as follows:
\bminiG{456}
  \label{eq:E456} 
  E_{4,5,6} \>=\> \frac43\left(1 \,+\,
  \frac{\sqrt{(1+3b^2)(1+3x^2)}}{2}\> \cos\left[\frac{\phi + 2k\pi}{3}
  \right]\right); \quad k = 0,\,1,\, 2\,, \qquad { }
\end{eqnarray}
where $\phi$ is given by
\begin{eqnarray}
\label{eq:R}
 \cos\phi=R\,, \qquad R = \frac{(1-9b^2)(1-9x^2)}
 {\left[(1+3b^2)(1+3x^2)\right]^{\frac32}}\,.
\emini
The last three eigenvectors are 
\begin{equation}
  \label{eq:V456}
  {{\cal{V}}_{4,5,6}} =  \left[
\begin{array}{c}
 \displaystyle -\frac {4}{N^2}\,(E_i \!-\! 1) \,b \\ [2ex]
 \displaystyle -\frac {N^{2}\! -\! 4}{N^2}\,(E_i\!-\!2) \,b \\ [2ex]
 \displaystyle \frac1{N^2-1}\left[ \left( E_i-\frac{N\!-\!1}{N}\right)
 \left(E_i-\frac{N\!+\!1}{N}\right) -\frac{N^2 \!-\! 5}{N^2}\,
 b^2\right] \\[2ex] \displaystyle \frac {4}{N^2} \,b^{2}\\ [2ex]
 \displaystyle \frac{N}{N+1}\left[ \frac{N\!+\!2}{2\,N} (E_i\!-\!2)
 \left(E_i-\frac{N\!+\!1}{N}\right) - {2\,b^2}\right] \\ [2ex]
 \displaystyle \frac{N}{N-1}\left[ \frac{N\!-\!2}{2\,N} (E_i\!-\!2)
 \left(E_i-\frac{N\!-\!1}{N}\right) - {2\,b^2}\right]
\end{array}
 \right] ,
\end{equation}
with $E_i$ the corresponding energy eigenvalue, $i=4,5,6$. 

Vectors $\cV_i$ are orthogonal with respect to the scalar product
defined by the metric tensor $W_{\al\be} = K_\al \delta_{\al\be}$, 
$$
 \left\langle \cV_i \right| W^{-1} \left| \cV_k\right\rangle \>=\> 0\,,
 \quad i\neq k\,.  
$$

\subsubsection{Strange symmetry}
We note an unexpected mysterious property of the equation
\eqref{eq:E13form} for the  eigenvalues of the soft
anomalous dimension matrix which is symmetric with respect to
\begin{equation} \label{eq:weird}
  b = \frac{T-U}{T+U} \quad \Longleftrightarrow \quad  x = \frac1N\,,
\end{equation} 
the transformation that interchanges parameters characterising 
external (scattering angle) and internal (colour group) degrees of freedom.

\subsection{Hard matrix element}
The colour structure of the hard gluon scattering matrix element can
be represented in terms of the $s$-, $t$- and $u$-channel projectors
 (see Appendix~\ref{AppProj})
\begin{equation}
  \label{eq:gpic}
  \cP_\a = \frac1N \quad \parbox{30pt}{ \begin{fmfgraph}(28,35)
      \fmfpen{thin}      
      \fmfincoming{l1,l2} \fmfoutgoing{r1,r2} 
      \fmf{gluon}{r2,u,r1}
      \fmf{gluon}{l1,d,l2}
      \fmf{gluon}{u,d}
      \fmfdot{d,u}
\end{fmfgraph}} , \quad  
  \cP_\a^{(t)} = \frac1N \quad \parbox{30pt}{ \begin{fmfgraph}(28,30)
      \fmfpen{thin}      
      \fmfincoming{l1,l2} \fmfoutgoing{r1,r2} 
      \fmf{gluon}{r2,u,l2}
      \fmf{gluon}{l1,d,r1}
      \fmffreeze
      \fmf{gluon}{d,u}
      \fmfdot{d,u}
\end{fmfgraph}}, \quad 
  \cP_\a^{(u)} = \frac1N \quad \parbox{40pt}{ \begin{fmfgraph}(38,30)
      \fmfpen{thin}      
      \fmfincoming{l1,l2} \fmfoutgoing{r1,r2} 
      \fmf{phantom}{l2,u}
      \fmf{phantom}{l1,d}
      \fmf{gluon}{r2,u}
      \fmf{gluon}{d,r1}
      \fmffreeze
      \fmf{gluon}{d,u}
      \fmf{gluon,right=0.25}{l2,d}
      \fmf{gluon,right=0.25}{u,l1}
      \fmfdot{d,u}
\end{fmfgraph}}.
\end{equation}
We have
\begin{equation}
  \label{eq:M0}
 M_0 = N\left(\,m_s\,\cP_\a +m_t\,\cP^{(t)}_\a +
m_u\,\cP^{(u)}_\a\,\right),  
\end{equation}
where we have included into $m_\lambda$ ($\lambda=s,t,u$) the one-gluon
exchange diagram in the $\lambda$-channel together with the piece of the
four-gluon vertex contribution that has the same colour structure.
The $t$- and $u$-channel projectors $P^{(t)}_\a$ and $P^{(u)}_\a$ can
be expressed in terms of the $s$-channel ones introduced in
\eqref{eq:projsum} (see Appendix~\ref{AppRepro}) as follows:
\begin{equation}
\begin{split}
  \cP^{(t)}_\a =  & \quad\> \frac{1}2\cP_\a + \cP_\sing +\frac{1}2\cP_\s
  -\frac1N  \cP_{\27}  +\frac1N  \cP_\0\, , \\
  \cP^{(u)}_\a =  & - \frac{1}2\cP_\a + \cP_\sing +\frac{1}2\cP_\s
  -\frac1N \cP_{\27} + \frac1N \cP_\0\,.
\end{split}
\end{equation}
We obtain
\begin{equation}
  \label{eq:M0-proj} 
  M_0 = N\left[\frac{M_a}{2} \cP_\a + M_s \left(\cP_\sing\!+\!\frac{1}2\cP_\s \!-\!\frac1N
  \cP_{\27}\!+\!
\frac1N \cP_\0 \right)\right],
\end{equation}
where $M_a$ and $M_s$ are, respectively, the parts of the matrix
element antisymmetric and symmetric with respect to exchange of gluons
in the $s$-channel:
\begin{equation}
  \label{eq:Mas}
  M_a \>=\> 2m_s+m_t-m_u\,, \quad M_s = m_t+m_u\,.
\end{equation}
It is worthwhile to notice that these amplitudes are separately gauge invariant.
Squaring the matrix element \eqref{eq:M0-proj} gives
\begin{equation}
  \label{eq:15}
\abs{ M_0}^2 \>=\> N^2 
  \left[\, \frac{M_a^2}{4}\cdot \cP_\a + M_s^2\cdot \left\{\cP_\sing +
  \frac14 \cP_\s + \frac1{N^2} \cP_\27 + \frac1{N^2} \cP_\0  \right\}
  \right].
\end{equation}
Here
\begin{equation}
  \label{eq:matrel}
\begin{split}
  M_a^2 &\equiv (2m_s\!+\!m_t\!-\!m_u)^2 = 9 -\frac{st}{u^2}
  -\frac{us}{t^2} - \frac{4 tu}{s^2} - \frac{3 s^2}{tu}, \\ M_s^2
  &\equiv \quad\> (m_t + m_u)^2 \quad =
  1-\frac{st}{u^2}-\frac{us}{t^2} +\frac{s^2}{tu} ,
\end{split}
\end{equation}
where we have used the known Lorentz matrix elements.

The total scattering cross section is proportional to the colour trace
of the squared matrix element \eqref{eq:15}:
\begin{equation}
 \sigma_0 \equiv \Tr\bigl( M_0^2 \bigr) = N^2(N^2\!-\!1)\frac{M_a^2 +
 3M_s^2}{4} = N^2(N^2\!-\!1)
\left\{3-\frac{tu}{s^2}-\frac{us}{t^2}-\frac{st}{u^2} \right\},
\end{equation}
where we have used the kinematical relation
$$
  \frac{s^2}{tu}+ \frac{t^2}{su}+ \frac{u^2}{st} \>=\> 3\,.
$$
Let us mention another elegant representation for the colour summed
cross section,
\begin{equation}
  \label{eq:m02}
   \sigma_0   = \frac{N^2}2 (N^2\!-\!1)
   \left[ (m_t\!+\!m_u)^2 + (m_u\!-\!m_s)^2 + (m_s\!+\!m_t)^2 \right].
\end{equation}
Here the first term in square brackets is given in \eqref{eq:matrel}
and the other two can be obtained from it by simple crossing, that is
by replacing $s \leftrightarrow t$ and $s \leftrightarrow u$,
respectively.

 \subsection{Squaring dressed matrix element}
 
Now we are in a position to construct the dressed matrix element
according to \eqref{eq:M-fine} and evaluate the cross
section. Expressing the eigenvectors \eqref{eq:V123}, \eqref{eq:V456} as
\begin{equation}
 \cV_\kappa = (Z\cdot \cP)_\kappa = \sum_\al Z_\kappa^\al\> \cP_\al\,,
 \qquad \cP_\al = (Z^{-1}\cdot \cV)_\al = \sum_\kappa
 \left(Z^{-1}\right)^\kappa_\al \cV_\kappa\,,
\end{equation}
for the evolution exponent we have
\begin{equation}
  e^{\Gamma\tau} M_0 \>= \> \sum_\beta m^\beta(\tau) \cP_\beta\,,
\end{equation}
where we have introduced 
\begin{equation}
m^\beta(\tau) \>=\> \sum_\al m^{(0)\al} \> \sum_\kappa
(Z^{-1})_{\alpha}^{\kappa}\cdot e^{-N(T+U)\tau\, E_\kappa} \cdot
Z_{\kappa}^{\beta}.
\end{equation}
The soft factor $\cS_X$ becomes
\begin{equation} \label{eq:Y}
\begin{split}
  \cS_X(\tau) \>&=\> \sigma_0^{-1} \Tr \left( M^\dagger(\tau)\cdot
  M(\tau)\right) = \sigma_0^{-1} \sum_\beta \abs{m^\beta(\tau)}^2 \cdot
  K_\beta\,.
\end{split}
\end{equation}

\subsection{Special cases}
 Now we turn to the discussion of special cases in which the answer is
 relatively simple and can be given explicitly.
 
\subsubsection{Scattering at $90^{\rm o}$}
   
Consider first the simple case of $b\!=\!0$ ($t=u$) which corresponds
to $90$ degree scattering.  Here $\cQ$ is diagonal so that the
$s$-channel projectors $\cP_{\al}$ become eigenvectors whose
eigenvalues are just the corresponding diagonal elements of $\cQ$:
\begin{eqnarray}
  \label{eq:E123456-b=0} E_\kappa &=& \left\{ 1,\> \frac32,\>
  \frac32,\> 2,\> \frac{N\!-\!1}{N},\> \frac{N\!+\!1}{N}
\right\} ,\\ 
\cV_\kappa &\propto& 
 \left\{ \cP_\10,\> \cP_\s+\cP_a,\> \cP_\s-\cP_a,\> \cP_\sing,\>
 \cP_\27,\> \cP_\0 \right\} .
\end{eqnarray}
To present the answer for the soft factor $\cS_X$ it is convenient to 
define the suppression factors
\begin{equation}
     \chi_t(\tau) \>=\> \exp\left\{-2N\tau
     \cdot\ln\frac{s}{-t}\right\}, \quad \chi_u(\tau) \>=\>
     \exp\left\{-2N\tau \cdot\ln\frac{s}{-u}\right\}.
\end{equation}
In $90^o$ scattering kinematics we have $\Re T=\Re U= \ln 2$ and
\begin{equation}
     \chi_t=\chi_u \>=\> \chi(\tau)\>=\> \exp\left\{-2N\tau\ln2\right\},
\end{equation}
and we get
\begin{equation} \label{eq:Sb0}
\cS_X(\tau) = \frac{\chi^2}{3}\left[\,\frac{4}{N^2\!-\!1}\, \chi^{2} + { \chi}
+ \frac{N\!-\!3}{N\!-\!1}\, \chi^{\frac2N} + \frac{N\!+\!3}{N\!+\!1}\,
\chi^{-\frac2N} \,\right].
\end{equation}
We check that for $\tau=0$, $\chi=1$, \eqref{eq:Sb0} gives indeed
$\cS_X(0)=1$ as it should. 

The cubic equation \eqref{eq:R} trivialises and can be solved
explicitly for $b\!=\!0$:
\begin{equation}
  \label{eq:b0}
\begin{split}
  & \cos\phi = \frac{1-9x^2}{(1+3x^2)^{3/2}},
  \quad\Longrightarrow\quad \cos\frac{\phi}{3} =
  \frac{1}{\sqrt{1+3x^2}}\,, \\ & \cos\left(\frac{\phi}{3} \pm
  \frac{2\pi}{3}\right) = -\frac{1\>\pm 3x}{2\sqrt{1+3x^2}} \,.
  \end{split}
\end{equation}
Substituting \eqref{eq:b0} into \eqref{eq:E456} gives the last three
energy levels in \eqref{eq:E123456-b=0}.

 \subsubsection{$N\to\infty$ limit}

By virtue of the weird symmetry \eqref{eq:weird}, the large-$N$ limit
($x\to0$) is related with the $90^o$ scattering case considered above:
the $N$-dependent energy levels 4, 5, 6 can be obtained from
\eqref{eq:E123456-b=0} simply by replacing $x\to b$,
\begin{equation}
  \label{eq:x0} 
  \cos\frac{\phi}{3} = \frac{1}{\sqrt{1+3b^2}}\,, \qquad
  \cos\left(\frac{\phi}{3} \pm \frac{2\pi}{3}\right) \>=\>
  -\frac{1\>\pm 3b}{2\sqrt{1+3b^2}} \,.
\end{equation}
The energy levels are as follows: 
$$ 
 E_1= 1, \quad E_2= \frac{3-b}{2},
 \quad E_3= \frac{3+b}{2}, \quad E_4= 2, \quad E_5= 1-b, \quad E_6=
1+b\,.  
$$ 
The weird symmetry does not extend upon the eigenvectors so that they
have to be derived anew:
\begin{equation}
\label{eq:Vx0}
\cV_{1, \ldots 6} \>=\>   
\left[ \begin {array}{r} 0\\\noalign{\medskip}0\\\noalign{\medskip}0
\\\noalign{\medskip}0\\\noalign{\medskip}1\\\noalign{\medskip}-1
\end {array} \right] 
\left[ \begin {array}{r} {1+b}\\\noalign{\medskip} -2b
\\\noalign{\medskip}0\\\noalign{\medskip}  {1+b}
\\\noalign{\medskip}-b\\\noalign{\medskip}-b \end {array} \right] 
\left[ \begin {array}{r} {1-b}\\\noalign{\medskip} 2b
\\\noalign{\medskip}0\\\noalign{\medskip} {-1+b}
\\\noalign{\medskip}-b\\\noalign{\medskip}-b\end {array} \right] 
 \left[ \begin {array}{r} -4(1\!-\!{b}^{2}) \\\noalign{\medskip}-8
 {b^3}
\\\noalign{\medskip} (1\!-\!{b}^{2})^{2}
\\\noalign{\medskip} 4 (1\!-\!{b}^{2})
\\\noalign{\medskip} 2b^2(1\!+\!b^2) \\\noalign{\medskip} 2b^2(1\!+\!b^2)
\end {array} \right] 
 \left[ \begin {array}{r} 0\\\noalign{\medskip}2\\\noalign{\medskip}0
\\\noalign{\medskip}0\\\noalign{\medskip}1\\\noalign{\medskip}1
\end {array} \right] 
 \left[ \begin {array}{r} 0\\\noalign{\medskip}-2\\\noalign{\medskip}0
\\\noalign{\medskip}0\\\noalign{\medskip}1\\\noalign{\medskip}1
\end {array} \right]. 
\end{equation}
The soft factor becomes
\begin{equation}
\label{eq:x0ans}
\begin{split}
  \cS_X \>&\simeq\> 
 \frac{\chi_t \, \chi_u}{2\,(M_a^2+3M_s^2)} \left[\, {4\,M_s^2} +
    {(M_a\!-\!M_s)^2}\,\chi_t +   {(M_a\!+\!M_s)^2}\,\chi_u\,\right].
\end{split} 
\end{equation}

\subsubsection{Regge limit}

In the case $b\to \pm 1$ (small angle scattering) \eqref{eq:R} can be
solved explicitly too, for arbitrary~$N$:
\begin{equation}
  \label{eq:b1}
\begin{split}
  & \cos\phi = -\frac{1-9x^2}{(1+3x^2)^{3/2}},
  \quad\Longrightarrow\quad \cos\frac{\phi}{3} =
  \frac{-1}{\sqrt{1+3x^2}}\,, \\ & \cos\left(\frac{\phi}{3} \pm
  \frac{2\pi}{3}\right) = \frac{1\>\mp 3x}{2\sqrt{1+3x^2}} \,.
  \end{split}
\end{equation}
Substituting \eqref{eq:b1} into \eqref{eq:E456} and invoking
\eqref{eq:E123} produces 
the set of eigenvalues
\begin{equation}
  \label{eq:Eallb1}
\{E_\kappa\} \>=\>  
\left\{ 1,\> 1,\> 2;\> 0,\> 2(1\!-\!x), 2(1\!+\!x) \right\}.  
\end{equation}
In what follows we consider the forward scattering case, $b\to+1$.

For $|t|\ll s\simeq |u|$ we have $ T\simeq \ln\frac{s}{-t} \gg
|U|\simeq \pi$. Neglecting the finite phase, in the logarithmic
approximation in $T$ the soft matrix \eqref{eq:cSans} becomes diagonal
in the $t$-channel basis
\begin{equation} F_X(\tau) = e^{\tau\Gamma} \>\simeq\>  K_{st}\,
e^{-C_2\cdot \tau\, \ln ( {s}/{t} )} \, K_{ts} ,
\end{equation} 
where $C_2$ is the diagonal matrix of the Casimirs \eqref{eq:C2matr}
and the matrices $K$ are given in Appendix~\ref{AppRepro}. These
exponents describe {\em reggeization}\/ of six possible $t$-channel
colour states. Indeed, the energy levels $N\cdot E_\kappa$ in
\eqref{eq:Eallb1} equal the Casimir operators. Cast in the canonical
$t$-channel order \eqref{eq:basis}, the eigenvalues read 
$$ 
c_\al \>=\> \{N,\,2N,\,0,\,N,2(N\!+\!1),\,2(N\!-\!1)\} \>=\> N\cdot
\{E_1,\> E_3, \> E_4,\> E_2, \> E_6,\> E_5 \}. 
$$
The eigenvectors corresponding to the energies \eqref{eq:Eallb1}
become, accordingly, pure $t$-channel projector states,
\begin{equation}\begin{split}
  \cV_1 & =K_{st}\cdot\cP^{(t)}_\a,\qquad
  \cV_2=K_{st}\cdot\cP^{(t)}_\s, \qquad
  \cV_3  =K_{st}\cdot\cP^{(t)}_\10,\\
  \cV_4 & =K_{st}\cdot\cP^{(t)}_\sing, \qquad
  \cV_5  =K_{st}\cdot\cP^{(t)}_\0,\qquad \cV_6\,.
  =K_{st}\cdot\cP^{(t)}_\27,
\end{split}
\end{equation}
and are given, correspondingly, by the {\em columns}\/ \#\ 1, 4, 2, 3,
6 and 5 of the re-projection matrix $K_{ts}$ \eqref{App:Kts}. In our
case of the order $\as$ matrix element, \eqref{eq:M0-proj} in the
$t\to0$ limit reduces to {\em only one state}\/ namely, that of the
asymmetric $t$-channel octet. Indeed, for $M_s=M_a\simeq m_t$ we have
$$ 
 M_0 \>\simeq\> N\cdot \cP^{(t)}_\a\,, \qquad \cF_X \cdot M_0 =
 \left(\frac{s}{t}\right)^{-N\tau}\cdot M_0\,,
$$
giving 
\begin{equation}
\label{eq:b1ans}
    S(\tau)\>=\> \chi_t(\tau) \>=\>
    \left(\frac{s}{t}\right)^{-2N\tau},
\end{equation}
which exponent coincides with the (twice) Regge trajectory of the
gluon exchanged in the $t$-channel.

\subsubsection{$N=3$} 

In $SU(3)$ the representation $\0$ has zero weight, $K_\0=0$, and the
projector $\cP_\0$ does not contribute. The projector basis reduces to
five states: $\al=\{\a,\10,\sing,\s,\27 \} $. The reduced anomalous
dimension matrix $\cQ$ is obtained from \eqref{eq:Qnew} by setting
$N\!=\!3$ and removing the last row and column. The first three
eigenvalues $E_{1,2,3}$ are $N$-independent and given by
\eqref{eq:E123}. The other three $E_{4,5,6}$ are easy to obtain from
\eqref{eq:E456} where $\phi=\pi/2$. Dropping the eigenvalue $E_6=4/3$
attached to the fake $\0$ state, we have five energy levels
\begin{equation}
  \label{eq:EN3} 
  E_\kappa = \left\{1 , \frac{3-b}{2}, \frac{3+b}{2},
  \frac{4+2\sqrt{1+3b^2}}{3}, \frac{4-2\sqrt{1+3b^2}}{3}\right\} .
\end{equation}
The corresponding eigenvectors ${{\cal{V}}_{1,\ldots 5}} $ read
\begin{equation}
  \label{eq:VN3}
\left[ \begin{array}{c}
 \displaystyle  0  \\ [2ex]
 \displaystyle 1 \\ [2ex]
 \displaystyle 0  \\[2ex]
 \displaystyle  \frac {4b}{5} \\ [2ex]
 \displaystyle  -\frac{9b}{5} 
\end{array}  \right] 
\left[ \begin{array}{c}
 \displaystyle  b\!+\!1  \\ [2ex]
 \displaystyle -2b \\ [2ex]
 \displaystyle \frac{b}2  \\[2ex]
 \displaystyle  1 \!-\!\frac {3b}{5} \\ [2ex]
 \displaystyle  -\frac{9b}{10}  
\end{array}  \right] 
\left[ \begin{array}{c}
 \displaystyle  b\!-\!1  \\ [2ex]
 \displaystyle -2b \\ [2ex]
 \displaystyle -\frac{b}2  \\[2ex]
 \displaystyle  1 \!+\! \frac {3b}{5} \\ [2ex]
 \displaystyle  \frac{9b}{10}  
\end{array}  \right] 
\left[ \begin{array}{c}
 \displaystyle -\frac{4b\left( 1 \!+\!2 \sqrt{1\!+\!3b^2} \right)}{27}
 \\ [2ex] \displaystyle \frac{10b\left( 1 \!-\! \sqrt{1\!+\!3b^2}
 \right)}{27} \\ [2ex] \displaystyle \frac{1 \!+\! 
 \sqrt{1\!+\!3b^2}}{18} + \frac{b^2}{9}\\[2ex] \displaystyle \frac
 {4\,b^2}{9} \\ [2ex] \displaystyle \frac{5\left( 1 \!-\! 
 \sqrt{1\!+\!3b^2} \right) }{18} + \frac{2b^2}{3}
\end{array}  \right] 
\left[ \begin{array}{c}
 \displaystyle -\frac{4b\left( 1 \!-\!2 \sqrt{1\!+\!3b^2} \right)}{27}
 \\ [2ex] \displaystyle \frac{10b\left( 1 \!+\! \sqrt{1\!+\!3b^2}
 \right)}{27} \\ [2ex] \displaystyle \frac{1 \!-\! \sqrt{1\!+\!3b^2}
 }{18} + \frac{b^2}{9} \\[2ex] \displaystyle \frac {4\,b^2}{9} \\
 [2ex] \displaystyle \frac{5\left( 1 \!+\! \sqrt{1\!+\!3b^2}
 \right)}{18} \!+\! \frac{2b^2}{3}
\end{array}  \right] \nonumber
\end{equation}

\section{Conclusions}

In this paper we considered the soft SL factor $\cS_X$ defined in
\eqref{eq:cS} that enters the general representation \eqref{eq:Sig}
for two scale QCD observables in hadron--hadron collisions. Being
collinear safe, this factor is driven by emission of soft gluons at
large angles, see~\eqref{eq:largeang}. For {\em global}\/ observables the
problem reduces to the analysis of soft radiation off the primary hard
partons $p_i$ ($i=1,\ldots 4$) only, and essentially trivialises in
spite of remaining non-Abelian.  Indeed, such accompanying radiation
has classical nature described by eikonal currents and Coulomb phase
effects. Virtual and real {\em eikonal}\/ contributions due to such
gluons fully cancel in the phase space region $k_t<Q_0$, while the
Coulomb contributions reduce to an (infinite) colour matrix phase that
cancels in the distributions.
The net result is the virtual eikonal suppression, accompanied by finite
non-Abelian Coulomb phase, due to the complementary momentum region
$Q_0<k_t<Q$.

Our first result is the general simple expression for virtual dressing
of the scattering matrix element, in terms of colour charges (Casimir
operators) of the cross-channel colour exchanges, $T_t^2$ and $T_u^2$:
\begin{equation}
 \frac12 \int (dk)\> j^2(k) \>=\> \frac12\, \sum_{i=1}^4
 T_i^2\cdot\cR_i
 \>+\> T_t^2\cdot T + T_u^2\cdot U. \quad \biggl( T=\ln\frac{s}{t},\>\>
U=\ln\frac{s}{u} \biggr)
\end{equation}
Here $\cR_i$ are the (colour-trivial) ``radiators'' that accommodate
collinear singularities and participate in forming the collinear
factor in \eqref{eq:Sig}. The last two terms form the soft anomalous
dimension matrix $\Gamma$ that determines $\cS_X$ as a function of the
SL variable $\tau$~\eqref{eq:tau}. Our anomalous dimension differs
from the one introduced in~\cite{KOS,BCMN} by a piece proportional to
the unit matrix. In our approach, this piece is absorbed into
collinear parton radiators. It participates in determining the precise
scales of the DL form factors in \eqref{eq:Sig} in terms of a
combination of angular integrated soft dipoles $W_{ij}^{(\ell)}$ each
of which produces the same scale $Q^2=tu/s$, see \eqref{eq:Qtudef}.

The matrices $T_t^2$ and $T_u^2$ do not commute. We found it
convenient to work in the colour basis of $s$-channel projectors where
each of them can be easily found with use of the re-projection
matrices \eqref{eq:Kall},
\begin{equation}
\label{eq:key}
  T_t^2 \>=\> K_{st}\, C_2^{(t)}\, K_{ts}, \quad T_u^2 \>=\> K_{su}\,
  C_2^{(u)}\, K_{us},
\end{equation}  
with $C_2$ the diagonal matrix of Casimir operators of all irreducible
representations present in the $t$ ($u$) channel.

Using the $s$-channel language makes the treatment and understanding
of the results more transparent.  The graphical colour projection
technique presented in the Appendix allowed us to avoid using the
over-complete Chan--Paton basis and largely simplified the analysis. In
particular, the calculation of the key ingredients \eqref{eq:key} of
the anomalous dimension becomes very simple since knowing the
transformation matrices $K$ the problem reduces to the Casimirs (for
gluon--gluon scattering the Casimirs are given in \eqref{eq:C2matr}).

Another advantage of the representation for $\Gamma$ in
\eqref{eq:M-fine} in terms of cross-channel charges \eqref{eq:key} is
trivialisation of the analysis of the Regge behaviour. In the case
of small angle scattering one term dominates, $T\gg U$ (forward
scattering) or $U\gg T$ (backward), and the anomalous dimension $\Gam$
becomes diagonal in the corresponding channel so that the problem
becomes essentially Abelian. 
Resulting exponents are nothing but Regge trajectories of $t$-($u$-)
channel exchanges that are proportional to corresponding Casimirs.

As an example we considered in detail the case of gluon--gluon
scattering which was first treated by Kidonakis, Oderda and Sterman
in~\cite{KOS}. Its colour structure is sufficiently complex as the
problem involves in general six colour states (which reduce to five in
$SU(3)$).
% to make theorists who dared to discuss it go nuts.
We found a simple representation for arbitrary $N$ for the eigenvalues
of the matrix $\cQ$, related with $\Gam$ as
$\Gam\!=\!-N(T\!+\!U)\,\cQ$, with $T=\ln(s/|t|)-i\pi$, $U=\ln(s/|u|)-i\pi$.  
In our representation the three $N$-dependent energy levels \eqref{456} and
corresponding eigenvectors \eqref{eq:V456} are explicitly real
functions of $T/U$ (the property not easy to extract from \cite{KOS}).

We gave explicit solutions for the soft factor $\cS_X$ in a number of
special cases including large-$N$ \eqref{eq:x0ans} and Regge limits
\eqref{eq:b1ans}.

Finally, we observed that the cubic equation \eqref{eq:E13form} for
the $N$-dependent energy levels 4, 5, 6 of $\cQ$ possesses a weird
symmetry which interchanges internal (colour group) and external
(scattering angle) degrees of freedom:
\begin{equation} \label{eq:weird2}
  \frac{T+U}{T-U} \quad \Longleftrightarrow \quad  N\,.
\end{equation} 
In particular, this symmetry relates $90$-degree scattering, $T=U$,
with the large-$N$ limit of the theory.  Giving the complexity of the
expressions involved, such a symmetry being accidental looks highly
improbable. Its origin remain mysterious and may point at existence of
an enveloping theoretical context that correlates internal and
external variables (string theory?).

\appendix

\section{Two gluon states in $SU(N)$ \label{AppProj}}

To construct colour states of the two-gluon system in the $s$-channel
we draw a pictorial identity
\begin{equation}
\label{eq:A1}
\U_{aa'}^{bb'}\equiv\> 
\delta_{aa'}\,\delta^{bb'}\>=\>\>\> 
\parbox{35pt}{ \begin{fmfgraph*}(25,30)
      \fmfpen{thin}      
      \fmfincoming{i1,i2}
      \fmfoutgoing{o1,o2}
      \fmf{gluon}{o2,i2}
      \fmf{gluon}{i1,o1}
      \fmfv{label=$b$,label.angle=-160}{i2}
      \fmfv{label=$b'$,label.angle=-20}{o2}    
      \fmfv{label=$a$,label.angle= 160}{i1}   
      \fmfv{label=$a'$,label.angle= 20}{o1}    
\end{fmfgraph*}}
=\>\> 4
\parbox{50pt}{ \begin{fmfgraph}(50,30)
      \fmfpen{thin}      
      \fmfincoming{i1,i2}
      \fmfoutgoing{o1,o2}
      \fmf{gluon}{o2,v2} \fmf{gluon}{u2,i2}
      \fmf{gluon}{i1,u1} \fmf{gluon}{v1,o1}
      \fmf{fermion,tension=0.2,left=0.5}{v2,u2,v2}
      \fmf{fermion,tension=0.2,left=0.5}{v1,u1,v1}
\end{fmfgraph}} 
\end{equation}
\smallskip

\noindent
where $a,b$ ($a',b'$) are colour indices of incoming (outgoing)
gluons, and analyse an intermediate state consisting of two quarks and
two antiquarks.
Here we have used
\begin{equation}
\parbox{50pt}{ \begin{fmfgraph}(50,15)
      \fmfpen{thin}      
      \fmfincoming{i1}
      \fmfoutgoing{o1}
      \fmfv{label=$a$,label.angle= 160}{i1}   
      \fmfv{label=$a'$,label.angle= 20}{o1}    
      \fmf{gluon}{i1,u1} \fmf{gluon}{v1,o1}
      \fmf{fermion,tension=0.2,left=0.5}{v1,u1,v1}
\end{fmfgraph}} 
\>=\>\>\> \tr (t^at^{a'}) = \frac12\delta_{aa'} \,.
\end{equation}
By interchanging quark and antiquark lines we can construct four tensors
with a given symmetry with respect to quark and, separately, 
antiquark colour indices,
\begin{equation}\label{Pisum}
 \U \>=\>  \Pi^+_+ + \Pi^+_- + \Pi^-_+  + \Pi^-_-\,.
\end{equation}
We get
$$
\Pi^u_d= \frac14\left( \>\> 
\parbox{30pt}{ \begin{fmfgraph}(25,25)
      \fmfpen{thin}      
      \fmfincoming{i1,i2}
      \fmfoutgoing{o1,o2}
      \fmf{gluon}{o2,i2}
      \fmf{gluon}{i1,o1}
\end{fmfgraph}} 
+ ud 
\parbox{30pt}{ \begin{fmfgraph}(25,30)
      \fmfpen{thin}      
      \fmfincoming{i1,i2}
      \fmfoutgoing{o1,o2}
      \fmf{gluon}{o2,i1}
      \fmf{gluon}{o1,i2}
\end{fmfgraph}} 
\right)
+ u 
\parbox{60pt}{ \begin{fmfgraph}(50,30)
      \fmfpen{thin}      
      \fmfright{b4,r1,t4} 
      \fmftop{t1,t2,t0,t3,t4} \fmfbottom{b1,b2,b0,b3,b4}
      \fmf{gluon}{t2,t1}
      \fmf{gluon}{b1,b2}
      \fmf{fermion}{t3,t2}    \fmf{fermion}{b3,b2}
      \fmf{plain}{b3,t2}      \fmf{plain}{b2,t3}
      \fmf{gluon}{t4,t3}      \fmf{gluon}{b3,b4}
\end{fmfgraph}}
+d
\parbox{60pt}{ \begin{fmfgraph}(50,30)
      \fmfpen{thin}      
      \fmfright{b4,r1,t4} 
      \fmftop{t1,t2,t0,t3,t4} \fmfbottom{b1,b2,b0,b3,b4}
      \fmf{gluon}{t2,t1}
      \fmf{gluon}{b1,b2}
      \fmf{fermion}{t2,t3}    \fmf{fermion}{b2,b3}
      \fmf{plain}{b3,t2}      \fmf{plain}{b2,t3}
      \fmf{gluon}{t4,t3}      \fmf{gluon}{b3,b4}
\end{fmfgraph}}
$$
where $u,d = \pm$ label the symmetry with respect to two ``internal'' 
quarks and two antiquarks, respectively.
Introducing the notation
$$
\U = 
\parbox{30pt}{ \begin{fmfgraph}(25,25)
      \fmfpen{thin}      
      \fmfincoming{i1,i2}
      \fmfoutgoing{o1,o2}
      \fmf{gluon}{o2,i2}
      \fmf{gluon}{i1,o1}
\end{fmfgraph}}, \qquad \X=  
\parbox{30pt}{ \begin{fmfgraph}(25,30)
      \fmfpen{thin}      
      \fmfincoming{i1,i2}
      \fmfoutgoing{o1,o2}
      \fmf{gluon}{o2,i1}
      \fmf{gluon}{o1,i2}
\end{fmfgraph}} \quad \biggl(\> = \de^b_{a'}\de^{b'}_a \> \biggr) ,
$$
and
$$
\W_+ =
\parbox{60pt}{ \begin{fmfgraph}(50,30) \fmfkeep{Wplus}
      \fmfpen{thin}      
      \fmfright{b4,r1,t4} 
      \fmftop{t1,t2,t0,t3,t4} \fmfbottom{b1,b2,b0,b3,b4}
      \fmf{gluon}{t2,t1}
      \fmf{gluon}{b1,b2}
      \fmf{fermion}{t3,t2}    \fmf{fermion}{b3,b2}
      \fmf{plain}{b3,t2}      \fmf{plain}{b2,t3}
      \fmf{gluon}{t4,t3}      \fmf{gluon}{b3,b4}
\end{fmfgraph}}
, \qquad
\W_-=
\parbox{60pt}{ \begin{fmfgraph}(50,30) \fmfkeep{Wminus}
      \fmfpen{thin}      
      \fmfright{b4,r1,t4} 
      \fmftop{t1,t2,t0,t3,t4} \fmfbottom{b1,b2,b0,b3,b4}
      \fmf{gluon}{t2,t1}
      \fmf{gluon}{b1,b2}
      \fmf{fermion}{t2,t3}    \fmf{fermion}{b2,b3}
      \fmf{plain}{b3,t2}      \fmf{plain}{b2,t3}
      \fmf{gluon}{t4,t3}      \fmf{gluon}{b3,b4}
\end{fmfgraph}} \quad 
\biggl( \>=  \Tr (\,t^b \,t^{a'}\,t^a\,t^{b'})\> \biggr),
$$ 
we write down these four combinations~as
\begin{equation}
\label{Pipms}
\begin{split}
 \Pi^+_+ &= \quart(\U+\X) + \W_+ + \W_- \\
 \Pi^-_- &= \quart(\U+\X) - \W_+ - \W_- \\
 \Pi^+_- &= \quart(\U-\X) + \W_+ - \W_- \\
 \Pi^-_+ &= \quart(\U-\X) - \W_+ + \W_- \>.
\end{split}
\end{equation}
For example, $\Pi^+_+$ is symmetric under interchanging quarks
and antiquarks, $\Pi^+_-$ is quark-symmetric and
antisymmetric with respect to antiquarks, etc.

The sum of \eqref{Pipms} obviously reproduces  \eqref{Pisum}.
Observing that interchanging the {\em gluons}\/  
$a\leftrightarrow b$ we have $\U\leftrightarrow \X$ and 
$\W_+\leftrightarrow \W_-$,
we conclude that
\begin{quote}
 $\Pi^+_+$ and $\Pi^-_-$ are symmetric, \\[1mm]
 $\Pi^+_-$ and $\Pi^-_+$ are antisymmetric  
\end{quote}
with respect to interchanging the 
gluon indices $a,b$ (corresponding to $t \leftrightarrow u$).

 \subsection{Projectors} 
The simplest projectors --- those onto the singlet and (antisymmetric
an symmetric) octet states of two gluons --- are obtained by
connecting a quark and an antiquark line in \eqref{eq:A1} and can be
represented graphically as
\begin{eqnarray}
\label{Pgg1}
 \cP_\sing &=& \frac1{N^2-1}\>\>
\parbox{30pt}{ \begin{fmfgraph}(30,30) 
      \fmfstraight
      \fmfpen{thin}     
      \fmfincoming{i1,i2}  \fmfoutgoing{o1,o2}
      \fmf{gluon,right=0.4}{i1,i2}
      \fmf{gluon,right=0.4}{o2,o1}
\end{fmfgraph} }  
\quad\qquad \biggl( \>= \frac1{N^2-1}\, \de_a^b\,\de_{a'}^{b'} \> \biggr), 
 \\[2mm]
\label{Pgg8a}
 \cP_{\a} &=& \frac1N\>\>
\parbox{40pt}{ \begin{fmfgraph}(40,30) 
      \fmfstraight
      \fmfpen{thin}     
      \fmfincoming{i1,i2}  \fmfoutgoing{o1,o2}
      \fmf{gluon}{i1,v1,i2}
      \fmf{gluon}{o2,v2,o1}
      \fmf{gluon}{v2,v1}
      \fmfdot{v1,v2}
\end{fmfgraph} }  \qquad\qquad \biggl( \> = \frac1N\, if_{abc}\,if_{cb'a'} 
                     \>=\> \frac1N\, T^c_{ab}T^c_{b'a'} \biggr),  \\[2mm]
\label{Pgg8s}
 \cP_{\s} &=& \frac{N}{N^2-4}\>\>
\parbox{40pt}{ \begin{fmfgraph}(40,30) 
      \fmfstraight
      \fmfpen{thin}     
      \fmfincoming{i1,i2}  \fmfoutgoing{o1,o2}
      \fmf{gluon}{i1,v1,i2}
      \fmf{gluon}{o2,v2,o1}
      \fmf{gluon}{v2,v1}
 \fmfv{d.sh=pentagram,d.f=1,d.size=4thick}{v1,v2}
\end{fmfgraph} }  \qquad 
 \biggl( \>= \frac{N}{N^2-4}\, d_{abc}\,d_{a'b'c}\> \biggr), 
\end{eqnarray}
where the dot in \eqref{Pgg8a} marks the standard tree-gluon vertex
(group structure constant) $if_{abc}$ and the star in \eqref{Pgg8s}
stands for the symmetric $d_{abc}$ symbol.

Introducing the notation for $s$-channel ``multiplication'' of two
graphs $A$ and $B$,
$$
   \bigl( A\cdot B\bigr)^{bb'}_{aa'} \>=\> \sum_{c,d}\,
   A_{ac}^{bd}\,B^{db'}_{ca'}\>, 
$$ 
one obtains~\cite{bible}
\begin{equation} \label{bib1}
 \W_\pm\cdot\cP_\sing = -\frac1{4N} \cP_\sing \,, \quad
 \W_\pm\cdot\cP_\a = 0 \,, \quad
 \W_\pm\cdot\cP_\s = -\frac1{2N}\cP_\s \,.
\end{equation}
These relations help us to construct remaining irreducible
representations. To this end we subtract from the tensors $\Pi^u_d$
\eqref{Pipms} their projections onto the singlet and two octets,
\eqref{Pgg1}--\eqref{Pgg8s}, and derive the four higher projectors:
\begin{eqnarray}
\label{27++}
\cP_{27} &=& \Pi^+_+ -  \frac{N-2}{2N}\cP_{\s} - \frac{N-1}{2N}\cP_\sing \,; \\
\label{0--}
 \cP_{0} \> &=& \Pi^-_- -  \frac{N+2}{2N}\cP_{\s} 
 - \frac{N+1}{2N}\cP_\sing \,; \\[2mm]
\label{eq:1010} 
 P_{10}&=& \Pi^+_- - \half \cP_{\a} \,,\quad P_{\overline{10}} =
 \Pi^-_+ - \half \cP_{\a}\,.
\end{eqnarray} 
For our purposes, the irreducible decuplet and anti-decuplet
representations \eqref{eq:1010} can be handled as a single state,
$\cP_\10 = P_{10} + P_{\overline{10}}$.  Written in full, the higher
representation projectors read
\begin{eqnarray}
\label{Pgg1010}
 \cP_\10  &=& \frac12 (\U-\X) - \cP_\a\,;\\
\label{Pgg27} 
 \cP_\27 &=& \frac14(\U+\X) - \frac{N-2}{2N}\cP_\s -
 \frac{N-1}{2N}\cP_\sing + (\W_++\W_-)\,; \\
\label{Pgg0} 
 \cP_\0\> &=& \frac14(\U+\X) - \frac{N+2}{2N}\cP_\s -
 \frac{N+1}{2N}\cP_\sing - (\W_++\W_-)\,.
\end{eqnarray}
Making use of \eqref{bib1} and of the relations~\cite{bible}
\begin{eqnarray}
 && 16\, \W_\pm\!\cdot\! \W_\pm =
 \U\>-\>\frac{N^2-1}{N^2}\cP_\sing\>-\>\frac{N^2\!-\!4}{N^2}\cP_\s
 \>-\>\cP_\a\,, \\ && 16\, \W_\pm\!\cdot\! \W_\mp =
 \X\>-\>\frac{N^2-1}{N^2}\cP_\sing\>-\>
\frac{N^2\!-\!4}{N^2}\cP_\s \>+\>\cP_\a\,,
\end{eqnarray}
it is straightforward to verify that the operators
\eqref{Pgg1010}--\eqref{Pgg0} are indeed projectors, 
\begin{equation}
   \cP_\al \cdot \cP_\be \> = \> \cP_\al \,\delta_{\al\be}\,.
\end{equation}

\subsection{Casimir operators \label{AppSEx}}

The Casimir operators for the singlet, $c_\sing=0$, and octet states,
$c_\a=c_\s=N$, are known. To obtain $c_R$ for higher representations
$R=\10,\27,\0$ we construct the total colour charge of the two-gluon
state as a sum of four quark generators,
\begin{equation}
\label{eq:ATR}
    T_R^a \>=\> t^a_1 + t^a_2 +\bar{t}^a_{1'} +\bar{t}^a_{2'}\,,  
\end{equation}
where $(1,2)$ and $(1',2')$ are ``internal'' colour lines of quarks
and antiquarks inside two gluons, see~\eqref{eq:A1}.
$N^2$ colour states of a $qq$ pair (1,2) split into symmetric (\6) and
antisymmetric ($\ba3$) irreducible representations,
\bminiG{projects33}
\parbox{30pt}{\begin{fmfgraph}(20,30) 
      \fmfpen{thin}  \fmfstraight       
      \fmftop{t1,t2} \fmfbottom{b1,b2} 
      \fmf{fermion}{t1,t2}
      \fmf{fermion}{b1,b2}
\end{fmfgraph} } \equiv
  (3)\otimes (3)\>&=&\> P_{\6} \>+\> P_{\ba3} \>; \\[2mm]
\label{proj6}
  P_{\6}\>\>&=&\>\> \frac{1}{2}\left(\quad
\parbox{40pt}{\begin{fmfgraph}(30,30) 
      \fmfpen{thin}  \fmfstraight       
      \fmftop{t1,t2} \fmfbottom{b1,b2} 
      \fmf{fermion}{t1,t2}
      \fmf{fermion}{b1,b2}
\end{fmfgraph} } 
+ 
\parbox{40pt}{ \begin{fmfgraph}(30,30)
      \fmfpen{thin} \fmfstraight 
      \fmftop{v1,v4}
      \fmfbottom{v5,v8}
      \fmf{fermion}{v1,v8}  
      \fmf{plain}{v5,v}\fmf{fermion}{v,v4}  
 \end{fmfgraph}}  \right) \\[4mm]
\label{projb3}
  P_{\ba3}\>\>&=&\>\> \frac12\left(\quad
\parbox{40pt}{\begin{fmfgraph}(30,30) 
      \fmfpen{thin}  \fmfstraight       
      \fmftop{t1,t2} \fmfbottom{b1,b2} 
      \fmf{fermion}{t1,t2}
      \fmf{fermion}{b1,b2}
\end{fmfgraph} } 
-
\parbox{40pt}{ \begin{fmfgraph}(30,30)
      \fmfpen{thin} \fmfstraight 
      \fmftop{v1,v4}
      \fmfbottom{v5,v8}
      \fmf{fermion}{v1,v8}  
      \fmf{plain}{v5,v}\fmf{fermion}{v,v4}  
 \end{fmfgraph}}   \right) . 
\qquad { }
\emini
A $\qq$ pair ($(1,1'), (1,2'), (2,1'), (2,2')$) can be in general in
the colour singlet (\sing) and colour octet (\8) characterised by the
projectors
\bminiG{projects}
\parbox{30pt}{\begin{fmfgraph}(20,30) 
      \fmfpen{thin}  \fmfstraight       
      \fmftop{t1,t2} \fmfbottom{b1,b2} 
      \fmf{fermion}{t1,t2}
      \fmf{fermion}{b2,b1}
\end{fmfgraph} } \equiv
  (3)\otimes (\bar{3})\>\>&=&\>\> P_{1} \>+\> P_{8} \>; \\[2mm]
\label{proj1}
  P_{\sing}\>\>&=&\>\> \frac{1}{N}\quad
\parbox{30pt}{ \begin{fmfgraph}(30,25)
      \fmfpen{thin} \fmfstraight 
      \fmftop{v1,v4}
      \fmfbottom{v5,v8}
      \fmf{fermion,left=0.5}{v1,v5}  
      \fmf{fermion,left=0.5}{v8,v4}  
 \end{fmfgraph}} \\[2mm]
\label{proj8}
  P_{\8}\>\>&=&\>\> 2\quad
\parbox{70pt}{ \begin{fmfgraph}(70,25)
      \fmfpen{thin}     
      \fmfincoming{i1,i2}
      \fmfoutgoing{o1,o2}
      \fmf{fermion}{v1,i1}
      \fmf{fermion}{i2,v1}
      \fmf{gluon}{v2,v1}
      \fmf{fermion}{o1,v2}
      \fmf{fermion}{v2,o2}
      \fmfdot{v1,v2} 
\end{fmfgraph} }
\emini
Remark that \eqref{projects} is nothing but the graphic representation
of the Fierz identity.

Squaring $T_R^a$ in \eqref{eq:ATR} and exploiting the symmetry
properties of the representations
\eqref{27++}--\eqref{eq:1010} with respect to quarks and antiquarks we
arrive at
\begin{equation}
  c_R \equiv (T^a_R)^2 
  \>=\>  4C_F  + 8v_{\3\ba3}(\8) + 2\left[\,v_{\3\3}(R_q) 
  +v_{\ba3\ba3}(R_{\bar{q}})\,\right] ,
\end{equation}
where $v$ is the one-gluon ``exchange potential'' between fermions,
and $R_q$ ($R_{\bar{q}}$) marks the irreducible $SU(3)$ representation
of the two-quark (antiquark) system.

Since our representations are {\em traceless}, an internal quark and
an antiquark are always in the {\em octet}\/ state, 
$$ 
v_{\3\ba3} P_\8 = v_{\3\ba3}(\8) P_\8 = t^a\,\bar{t}^a\cdot 
P_\8 = -t^a\,{t}^a\cdot P_\8 = \frac1{2N}\, \cP_\8\,.  
$$ 
$R_q$ may be either $\6$ or $\ba3$ depending on the quark symmetry
($\b6$ and $\3$ for $R_{\bar{q}}$):
\begin{equation} \label{eq:qqpots} 
 v_{\3\3}(\6)= v_{\ba3\ba3}(\b6)=-
\frac{1-N}{2N} \,, \quad v_{\3\3}(\ba3)= v_{\ba3\ba3}(\3)= -
\frac{1+N}{2N}\,.  
\end{equation} 
The values of inter-quark potentials $v_{33}(\6)$ and $v_{33}(\ba3)$,
and $v_{3\bar{3}}(\8)$ between a quark and an antiquark are easy to
derive by projecting the Fierz identity \eqref{projects} in the
``rotated'' form,
\begin{equation} \label{eq:Fierz}
 \parbox{50pt}{ \begin{fmfgraph}(50,30)
      \fmfpen{thin}     
      \fmfincoming{i2,i1}
      \fmfoutgoing{o2,o1}
      \fmf{fermion}{i1,v1,o1}
      \fmf{fermion}{o2,v2,i2}
      \fmf{gluon,tension=0}{v2,v1}
      \fmfdot{v1,v2} 
\end{fmfgraph} }
 =\quad \frac12\quad 
\parbox{40pt}{ \begin{fmfgraph}(40,30)
      \fmfpen{thin} \fmfstraight 
      \fmftop{v1,v2,v3,v4}
      \fmfbottom{v5,v6,v7,v8}
      \fmf{vanilla}{v1,v2} \fmf{fermion}{v2,v6} \fmf{vanilla}{v6,v5}
      \fmf{vanilla}{v4,v3} \fmf{fermion}{v7,v3} \fmf{vanilla}{v7,v8}
      \fmf{phantom,tension=3}{v2,v3}  \fmf{phantom,tension=3}{v7,v6}  
\end{fmfgraph} }
\quad -\frac1{2N}\quad
\parbox{40pt}{ \begin{fmfgraph}(40,30)
      \fmfpen{thin} \fmfstraight 
      \fmftop{v1,v2,v3,v4}
      \fmfbottom{v5,v6,v7,v8}
      \fmf{vanilla}{v1,v2} \fmf{fermion}{v2,v3} \fmf{vanilla}{v3,v4}
      \fmf{vanilla}{v5,v6} \fmf{fermion}{v7,v6} \fmf{vanilla}{v7,v8}
\end{fmfgraph} }\qquad  ,
\end{equation}
onto the $s$-channel states \eqref{projects33} and \eqref{proj8}.
Thus we obtain
\begin{equation}
  c_R \>=\>  4\frac{N^2-1}{2N} + \frac{4}{N} + 2\left( \frac{u\,
      N-1}{2N} + \frac{ d\,  N-1}{2N}\right),  
\end{equation}
giving
\begin{equation}
 \label{eq:C270}
 \begin{split}
 c_\27 &  \>=\> 2(N+1),   \qquad (u=d=+)\\
 c_\0   &  \>=\> 2(N-1), \qquad (u=d=-) \\
 c_\10  & \>=\> 2N\,.   \qquad \qquad\>\> (u=-d)  
\end{split}
\end{equation} 

\subsection{$t$- and $u$-channel projectors in terms 
            of $s$-channel ones \label{AppRepro}}

$t$-channel gluon exchange between gluons ($s$-channel gluon exchange
potential) is proportional to the $t$-channel antisymmetric octet projector
$\cP^{(t)}_\a$ and has the following decomposition in $s$-channel colour
projectors
\begin{equation}
  \label{eq:14}
V  \equiv 
 \parbox{30pt}{ \begin{fmfgraph}(25,30)
      \fmfpen{thin}      
      \fmfincoming{i1,i2} \fmfoutgoing{o1,o2}
      \fmf{gluon}{o2,v2,i2}
      \fmf{gluon}{i1,v1,o1}
      \fmffreeze
      \fmf{gluon}{v2,v1}
 \fmfdot{v1,v2}    
\end{fmfgraph}} \!\!\!
= N\cP^{(t)}_\a =  \frac{N}{2}\cdot\cP_\a + 0\cdot \cP_\10 +   
 N\cdot \cP_\sing + \frac{N}{2}\cdot \cP_\s 
 + (-1) \cdot \cP_{\27} + 1\cdot \cP_{\0}.
\end{equation}
The coefficients in  \eqref{eq:14} are related with  the Casimir
operators of the representations involved, 
$c_\al= (T_1+T_2)^2$  and $T_1^2=T_2^2=N$, as 
\begin{equation}
 \label{eq:V2} 
 V \cdot \cP_\al \>\>=\>\> \frac{1}{2} \left[\, 2N -
 c_\al\,\right] \cdot \cP_\al\,,
\end{equation}
with $c_\al$ given in \eqref{eq:C2matr}.  The $u$-channel projector
$\cP_\a^{(u)}$ is obtained from \eqref{eq:14} by transposing the
outgoing gluons which amount to changing sign of the asymmetric
components $\cP_\a$ and $\cP_\10$.

The re-projection matrix $K_{ts}$ introduced in \eqref{eq:Kts} is given
by
\begin{equation}\label{App:Kts}
K_{ts} = \left( \begin{array}{rrrrrr}
 \frac12  & 0 & 1 & \frac12 & -\frac1N & \frac1N \\[1mm]
 0 & \frac12 & \frac{N^2-4}{2} & -1 & -\frac{N-2}{2N} & -\frac{N+2}{2N} \\[1mm]
 \frac{1}{N^2-1} & \frac{1}{N^2-1} 
&  \frac{1}{N^2-1} & \frac{1}{N^2-1} &
\frac{1}{N^2-1} & \frac{1}{N^2-1} \\[1mm]
 \frac12 & -\frac2{N^2-4} & 1 & \frac{N^2-12}{2(N^2-4)} &
\frac{1}{N+2} & -\frac{1}{N-2}  \\[1mm]
 -\frac{N(N+3)}{4(N+1)} & - \frac{N(N+3)}{4(N\!+\!1)(N\!+\!2)} 
& \frac{N^2(N+3)}{4(N+1)} &\frac{N^2(N+3)}{4(N\!+\!1)(N\!+\!2)} & 
\frac{N^2+N+2}{4(N\!+\!1)(N\!+\!2)} & \frac{N+3}{4(N+1)}  \\[1mm]
 \frac{N(N-3)}{4(N-1)} & - \frac{N(N-3)}{4(N\!-\!1)(N\!-\!2)} &
 \frac{N^2(N-3)}{4(N-1)} & -\frac{N^2(N-3)}{4(N\!-\!1)(N\!-\!2)} &  
\frac{N-3}{4(N-1)} & \frac{N^2-N+2}{4(N\!-\!1)(N\!-\!2)}   \\[1mm]
\end{array}  \right)
\end{equation}
After some reflection it is easy to understand why does 
the inverse matrix coincide with the direct one:  
\begin{equation}
 K_{st} \>=\> K_{ts} \,.
\end{equation}
It is also easy to construct the $u$-channel re-projection matrices
\eqref{eq:Kus} exploiting the symmetry of the $s$-channel projectors
under $t\leftrightarrow u$ transformation. Thus $K_{us}$ is obtained
by changing sign of the first two {\em columns}\/ of $K_{ts}$ and the
inverse matrix $K_{su}$ --- by changing sign of the first two {\em
rows}\/ of~$K_{ts}$.

\paragraph{Hints for deriving \eqref{App:Kts} without 
           spilling much blood \cite{bible}.}
\begin{itemize}
\item[$\cP^{(t)}_\a $:]  
is $1/N$ times the $s$-channel "gluon exchange potential"; see
\eqref{eq:14}.

\item[$\cP^{(t)}_\10 $:] Look at \eqref{Pgg1010} from $t$-channel perspective,
$$ 
 \cP_\10^{(t)} = \frac12 (\U^{(t)}-\X^{(t)}) - \cP_\a^{(t)}; \qquad
 \U^{(t)} \equiv (N^2\!-\!1)\cdot \cP_\sing, \quad \X^{(t)}\equiv \X .  
$$ 
Representation of the last term $\cP_\a^{(t)}$ we already know. The
cross we get from the representation of the unity, by exchanging
$t\leftrightarrow u$:
\begin{eqnarray*}
\U &=& \quad \cP_{\a} + \cP_\10 + \cP_\sing + \cP_\s + \cP_\27 + \cP_\0, \\
\X &=& \>-\cP_{\a} - \cP_\10 + \cP_\sing + \cP_\s + \cP_\27 + \cP_\0.
\end{eqnarray*}

\item[$\cP^{(t)}_\sing $:] $\equiv (N^2\!-\!1)^{-1}\cdot \U$. 

\item[$\cP^{(t)}_{\s} $:]  
Using pictorial representation for $if_{abc}$ and $d_{abc}$ symbols in
terms of quark loops,
\begin{equation}
i f_{abc}\,[d_{abc}] 
 =\>  2\left[ \quad
 \parbox{50pt}{ \begin{fmfgraph}(40,40)
      \fmfpen{thin}
      \fmftop{t1} \fmfbottom{b1,b2}
      \fmf{gluon}{b1,v1}  \fmf{gluon}{b2,v2}  \fmf{gluon}{v3,t1}
      \fmf{fermion,tension=0.5}{v1,v2,v3,v1}
\end{fmfgraph} }
\mp
 \parbox{50pt}{ \begin{fmfgraph}(40,40)
      \fmfpen{thin}
      \fmftop{t1} \fmfbottom{b1,b2}
      \fmf{gluon}{b1,v1}  \fmf{gluon}{b2,v2}  \fmf{gluon}{v3,t1}
      \fmf{fermion,tension=0.5}{v1,v3,v2,v1}
\end{fmfgraph} }
 \right],
\end{equation}
and the Fierz identity \eqref{eq:Fierz} to get rid of gluons
connecting quark triangles, it is straightforward to derive
\begin{equation} \label{App:B+B}
\frac14\left[   \>\>
   \parbox{30pt}{ \begin{fmfgraph}(25,30)        \fmfkeep{ddt}
      \fmfpen{thin}      
      \fmfincoming{i1,i2} \fmfoutgoing{o1,o2}
      \fmf{gluon}{o2,v2,i2}
      \fmf{gluon}{i1,v1,o1}
      \fmffreeze
      \fmf{gluon}{v2,v1}
  \fmfv{d.sh=pentagram,d.f=1,d.size=4thick}{v1,v2} 
\end{fmfgraph}}\>\> + \>\>
  \parbox{30pt}{ \begin{fmfgraph}(25,30)                \fmfkeep{fft}
      \fmfpen{thin}      
      \fmfincoming{i1,i2} \fmfoutgoing{o1,o2}
      \fmf{gluon}{o2,v2,i2}
      \fmf{gluon}{i1,v1,o1}
      \fmffreeze
      \fmf{gluon}{v2,v1}
   \fmfdot{v1,v2} 
\end{fmfgraph}}
 \right] \>\>+\>\>   \frac1{2N} \> 
 \parbox{30pt}{ \begin{fmfgraph}(25,25)           \fmfkeep{Us}
      \fmfpen{thin}      
      \fmfincoming{i1,i2}
      \fmfoutgoing{o1,o2}
      \fmf{gluon}{o2,i2}
      \fmf{gluon}{i1,o1}
\end{fmfgraph}}  \quad = \quad  \B_+ + \B_-  , 
\end{equation}
where $\B_\pm$ stand for {\em quark boxes}\/ with positive (negative)
direction of the fermion line: $\B_+=\Tr [t^{a_2}t^{a_1}t^{a_3}t^{a_4}]$,
$\B_-=\Tr[t^{a_1}t^{a_2}t^{a_4}t^{a_3}]$. Since the boxes are
rotationally invariant, we can {\em rotate}\/ the l.h.s.\ by $90^{\rm
o}$ and write the same expression in terms of the $s$-channel
operators:
\begin{equation}\label{App:B+B2}
\B_+ +\B_-  \>\equiv\> \B_+^{(t)} +\B_-^{(t)} 
 \>=\> \frac{N}{4} \cP_\a + \frac{N^2-1}{2N}\cP_\sing
 +\frac{N^2-4}{4N} \cP_\s.
\end{equation}
Equating \eqref{App:B+B}=\eqref{App:B+B2}, and already knowing
$\cP_\a^{(t)}$, suffices to get hold of $\cP_\s^{(t)}$.

\item[$\cP^{(t)}_{\27,\0} $:]

For the $t$-channel \27\ and \0\ projectors we use "rotated"
definitions \eqref{Pgg27} and \eqref{Pgg0}.  In these expressions we
know all the elements but the sum of rotated crossed boxes $W_+^{(t)}
+ W_-^{(t)}$. Flipping \eqref{App:B+B} around the top (that is,
exchanging $1\leftrightarrow 3$) we have
\begin{equation}  \label{App:W+W}
\frac14\left[   \>\>
   \parbox{30pt}{\fmfreuse{ddt} } \>\> - \>\>
   \parbox{30pt}{\fmfreuse{fft}} \right] \>\>+\>\> \frac1{2N} \>
   \parbox{30pt}{\fmfreuse{Us}} \quad = \quad \W_+ \>+\> \W_- .
\end{equation}
"Rotating" \eqref{App:W+W} by $90^{\rm o}$ we obtain 
\begin{equation}
\W_+^{(t)} +\W_-^{(t)}  \>=\> -\frac{N}{4} \cP_\a 
+ \frac{N^2-1}{2N}\cP_\sing +\frac{N^2-4}{4N} \cP_\s.
\end{equation}
\end{itemize}
Finally, it does not hurt to check the $t$-channel completeness
relation: $\sum_\rho \cP^{(t)}_\rho = \U^{(t)}$.

\section{Symmetric basis}

The soft anomalous dimension matrix $\cQ$ can be made symmetric by
applying the metric operation which consists of multiplying the
columns and dividing the rows by the square root of the dimension of
the representation.  That is,
\begin{eqnarray*}
  \Gam^{\rm symm} =  \cM^{-1}\,\Gam\,\cM\,,\quad
    \cM_{\al\be} = \sqrt{K_\al} \,\delta_{\al\be}\,,§
\end{eqnarray*}
with $K_{\al}$ given in \eqref{eq:Fs}.
In the symmetrised form the $\cQ$ matrix reads
\begin{equation}
  \label{eq:Qggsym}
%{\bf 2}\, N\,\Gam = 2N(T+U)\left( 
%\Gam =(T+U)\left( 
\cQ^{(s)} =\left( 
\begin{array}{cc|cccc}
\frac32     & 0 & \quad -\frac{2b}{U_1D_1} \quad   & -\frac{b}2 & 
- \frac{b\, U_3}{N\,U_1}  & 
-\frac{b\,D_3}{N\,D_1}\\[2ex] 0 & 1 & 0 & -\frac{b\,\sqrt{2}}{U_2D_2}
\quad & -\frac{b\,U_1D_2U_3}{\sqrt{2}N\,U_2} \quad &
-\frac{b\,D_1U_2D_3}{\sqrt{2}N\,D_2} \\[1ex] \hline -\frac{2b}{U_1D_1}
& 0 & 2 & 0 & 0 & 0 \\[2ex] -\frac{b}{2} & -\frac{b\,\sqrt{2}}{U_2D_2}
& 0 & \frac32 & 0 & 0 \\[2ex] -\frac{b\, U_3}{N\,U_1} &
-\frac{b\,U_1D_2U_3}{\sqrt{2}N\,U_2} &0 &0 & \frac{N-1}{N} & 0 \\[2ex]
-\frac{b\,D_3}{N\,D_1} \quad & -\frac{b\,D_1U_2D_3}{\sqrt{2}N\,D_2} &0
&0 & 0& \frac{N+1}{N}
\end{array}
\right)
\end{equation}
where we used shorthand notation
 $$
  U_k = \sqrt{N+k}\,, \quad D_k=\sqrt{N-k}\,.
 $$

\end{fmffile}

\begin{thebibliography}{99}

\bibitem{DS} 
M.\ Dasgupta and G.P.\ Salam, 
%``Resummation of non-global QCD observables,''
\plb{512}{2001}{323} [hep-ph/0104277]; \\
 M.\ Dasgupta and G.P.\ Salam, 
%``Accounting for coherence in interjet E(t) flow: A case study,''
\jhep{03}{2002}{017} [hep-ph/0203009]

\bibitem{NONGLOBlargeN} 
%``Away-from-jet energy flow,''
%JHEP {\bf 0208} (2002) 006 [hep-ph/0206076].
A.\ Banfi, G.\ Marchesini and G.\ Smye, 
\jhep{08}{2002}{006} [hep-ph/0206076]; \\
 Yu.L.\ Dokshitzer and G.\ Marchesini,  
% ON LARGE ANGLE MULTIPLE GLUON RADIATION.
\jhep{03}{2003}{040}   [hep-ph/0303101]
\bibitem{BS}
  J.\ Botts and G.\ Sterman, \npb{325}{1989}{62}
\bibitem{KOS}
  N.\ Kidonakis and G.\ Sterman, \plb{387}{1996}{867},
  \npb{505}{1997}{321} [hep-ph/9705234];\\
  N.\ Kidonakis, G.\ Oderda and G.\ Sterman, \npb{531}{1998}{365}
  [hep-ph/9803241];\\
  G.\ Oderda \pr{61}{2000}{014004} [hep-ph/9903240]\\
\bibitem{BCMN}
R.\ Bonciani, S.\ Catani, M.\ Mangano and P.\ Nason, \plb{575}{2003}{268} 
[hep-ph/0307035] \\
\bibitem{ASZ} 
A.\ Banfi, G.P.\ Salam and G.\ Zanderighi, 
% GENERALIZED RESUMMATION OF QCD FINAL STATE OBSERVABLES.
\plb{584}{2004}{298}
%% RESUMMED EVENT SHAPES AT HADRON - HADRON COLLIDERS.
% \jhep{08}{2004}{062}2004 [hep-ph/0407287];
\bibitem{DMlett}  
Yu.L.~Dokshitzer and G.~Marchesini [hep-ph/0508130]
\bibitem{QVscale}  % many collinear refs
% Thrust in E+E-
S.~Catani, L.~Trentadue, G.~Turnock and B.R.~Webber,
\npb{407}{1993}{3}; \\
% C-PARAMETER IN E+ E- ANNIHILATION.
S. Catani and B.R. Webber, \plb{427}{1998}{377}
[hep-ph/9801350];\\ 
%The Immortal Unappreciated broadening paper
Yu.L.~Dokshitzer, A.~Lucenti, G.~Marchesini and G.~P.~Salam,
%``On the {QCD} analysis of jet broadening,''
\jhep{01}{1998}{011}[hep-ph/9801324]
\bibitem{DKT3}  
 % COLLECTIVE QCD EFFECTS IN THE STRUCTURE OF FINAL MULTI - HADRON STATES. 
Yu.L.\ Dokshitzer, S.I.\ Troian and V.A.\ Khoze,
 {\it Sov.\ J.\ Nucl.\ Phys.\ }{\bf 46} (1987) 712,   
%, Yad.Fiz.46:1220-1232,1987
% MULTIPLE HADROPRODUCTION IN HARD PROCESSES WITH NONTRIVIAL TOPOLOGY
% By Yu.L.\ Dokshitzer, S.I.\ Troian and V.A.\ Khoze, 
% {\it Sov.\ J.\ Nucl.\ Phys.\ }{\bf 47} (1988) 881, % -888 % Yad.Fiz.47:1384-1396,1988
{\em ibid.}\/ {\bf 47} (1988) 881

\bibitem{Higgs} Yu.L.\ Dokshitzer, V.A. Khoze and S.I.\ Troian, {\em
    in}\/ Proceedings of the 6$^{\mbox{\scriptsize th}}$ Int.\ 
  Conference on Physics in Collisions 1986, ed.\ M.\ Derrick (World
  Scientific, Singapore, 1987), p.~365

\bibitem{bible} Yu.L.\ Dokshitzer, ``QCD for Beginners'' (unpublished).

\end{thebibliography}
\end{document}